\documentclass[twocolumn,showpacs,preprintnumbers,amsmath,amssymb]{revtex4}

\usepackage{graphicx}
\usepackage{dcolumn}
\usepackage{bm}
\usepackage{epsfig}
\usepackage{subfigure}

\begin{document}

\title{Quasi-cycles in a spatial predator-prey model}

\author{Carlos A. Lugo} 
\author{Alan J. McKane}
\email{alan.mckane@manchester.ac.uk} 

\affiliation{
Theoretical Physics Group, School of Physics and Astronomy\\ 
University of Manchester, Manchester M13 9PL, UK \\}

\date{\today}

\begin{abstract}
We show that spatial models of simple predator-prey interactions 
predict that predator and prey numbers oscillate in time and space. These 
oscillations are not seen in the deterministic versions of the models, but 
are due to stochastic fluctuations about the time-independent solutions of 
the deterministic equations which are amplified due to the existence of a 
resonance. We calculate the power spectra of the fluctuations analytically 
and show that they agree well with results obtained from stochastic 
simulations. This work extends the analysis of these quasi-cycles from that 
previously developed for well-mixed systems to spatial systems, and shows 
that the ideas and methods used for non-spatial models naturally generalize 
to the spatial case.
\end{abstract}


\maketitle

\section{Introduction}
\label{intro}
The standard paradigm of condensed matter physics involves the interaction 
of discrete entities (for example atoms, molecules or spins) positioned on
the sites of a regular lattice which when viewed at the macroscale can be
described by a differential equation after coarse-graining. 
This type of structure is not unique to 
physics; there are many other systems which consist of a large number of 
discrete entities which interact with each other in a simple way, but which
when viewed macroscopically show complex behavior. What is different,
however, is that physicists stress the relationships between models of the
same phenomena constructed at different scales, for instance by deriving 
macroscopic models from those defined at the microscale. Here we will be 
interested in modeling species in an ecological system where the interaction 
between individuals of those species is of the predator-prey type. Although 
both ``microscopic models'' --- individual based models (IBMs) defined on a 
two-dimensional lattice for example, and ``macroscopic models'' such as 
reaction-diffusion equations, have been extensively studied\,\cite{gri96}, the 
derivation of the latter from the former has received very little 
attention. Thus it is not obvious a priori if the results from the two 
different approaches can be meaningfully compared or if the macroscopic 
description misses some important features which are present in the IBM.

In this paper we will build on some earlier work\,\cite{AJMaTJN04} that 
introduced a methodology which began from a specific IBM and derived the 
corresponding model which holds at the macroscopic, or population, level.
The latter was called the population level model (PLM) and the former sometimes
called the individual level model (ILM), rather than the IBM, by analogy. 
There is another reason for using the term ILM in place of IBM. The nature
of the ``microscopic model'' can vary considerably. At one extreme are models
where the constituents each have individual characteristics. They may have 
an age, sex, be hungry at a given time, and so on. These are essentially agent 
based models\,\cite{VG,FS03}. At the other extreme are very simple 
``physical'' models, such as lattice gas models \cite{AP04}, where the 
analogies to physical processes take a primary role. The term IBM is  
frequently used for the former agent based models. Our starting point will be
somewhere between these two extremes. We model the individuals as entities
which may be born or die, may migrate to neighboring sites on the lattice in
a single time interval and when on the same lattice site may interact with 
each other if one is a prey species of the other. Thus the individuals act
as chemical species which have given interaction rules. There are several 
advantages with this formulation. Firstly, it corresponds most directly in
terms of properties of the constituents to PLMs such as the Volterra equations.
Secondly, more properties of individuals can be included if required,
taking the model more towards the agent-based IBMs mentioned above. Thirdly,
it allows the stochastic nature of birth, death, predator-prey and 
migrationary processes to be naturally included into the model. Whereas most 
stochastic models have been simulated directly, we prefer to formulate them 
as a master equation, and use the system-size expansion\,\cite{NGvK92} to 
derive the form they take when the system size is large.

The aim of this paper is then to investigate the nature of the PLM model both
at the macroscopic or mean-field level --- which is deterministic --- and at 
what might be described as the mesoscopic scale where stochastic effects are 
still important, but where the discrete nature of the constituents has been 
lost. The former is interesting because it is not clear that the model derived
in this way will coincide with those appearing in the textbooks on the 
subject\,\cite{WSCG98,NRMN03,TGH,JDM89}, but also because of the types of 
collective patterns frequently displayed by these systems, which often 
resemble those observed when studying physical and chemical systems. The 
latter is interesting because it has been found that in simple predator-prey 
models (without spatial effects being included) large predator-prey cycles are 
present in the stochastic model, which are lost at the deterministic 
level\,\cite{AJMaTJN05}. More specifically, the discrete nature of the 
individuals results in a demographic stochasticity at the mesoscale which
acts as a driving force and creates a resonance effect, turning small 
cyclic fluctuations into large cycles called quasi-cycles\,\cite{NRMN03}.
Here we investigate the nature of this phenomenon in a model where spatial 
effects are included. The ordinary differential equations of the Volterra 
type will now be replaced by partial differential equations of the 
reaction-diffusion type, and the two coupled Langevin equations of 
\cite{AJMaTJN05} will be replaced by two coupled partial differential 
equations with additive noise. 

The paper is organized as follows. In Section \ref{model}, the model alluded 
to above is introduced and formulated as a master equation. This is followed 
in Section III by a discussion of the deterministic limit of the equation, a 
linear stability analysis of the stationary solution of this equation, 
and the linear noise correction to the deterministic equation. In Section IV
a Fourier analysis of the linear stochastic differential equations is carried 
out which yields power spectra which characterize the nature of the spatial
and temporal predator-prey cycles, with the analytic results being compared 
to the results of computer simulations. There are two Appendices containing
mathematical details. The first describes the application of the system-size
expansion to the master equation and the second contains the Fourier analysis
of the linear stochastic differential equation.

\section{Model}
\label{model}
The system we will be interested in consists of individuals of species $A$ 
who are predators of individuals belonging to the prey species $B$. We 
assume that they inhabit patches, labeled by $i=1,\ldots,\Omega$, which are
situated at the sites of a $d-$dimensional hypercubic lattice. Of course, for
applications we are interested in the case of a square lattice in two 
dimensions, but we prefer to work with general $d$. One reason is that it is
not any more complicated doing so, another is because our stochastic 
simulations have been carried out in $d=1$ in order to achieve higher accuracy.
Each patch possesses a finite carrying capacity, $N$, which is the maximum 
number of individuals allowed per site. The number of predators and prey in 
patch $i$ will be denoted by $n_i$ and $m_i$ respectively. There will therefore
be $(N-n_i-m_i)$ empty or vacant ``spaces'', $E$, in patch $i$. These are 
necessary to allow the number of $A$ and $B$ individuals in patch $i$ to 
independently vary with time. Further background to the modeling procedure 
is given in Ref.~\cite{AJMaTJN04}, where it has been applied to competition 
between two species. 

As discussed in Section \ref{intro}, we assume that the constituents $A$, $B$
and $E$ react together at given rates. The reactions corresponding to birth,
death and predation are assumed to be local, that is, only involve individuals
at a particular site. They will therefore be identical to those invoked in
the predator-prey model without spatial structure\,\cite{AJMaTJN05}, and 
since these have been shown to lead to the Volterra equations in the
deterministic limit, we will adopt them here:
\begin{eqnarray}
B_{i}E_{i} & \xrightarrow{{b}} & B_{i}B_{i}\,, \label{spr1} \\ 
A_{i}B_{i} \xrightarrow{{p_{1}}} A_{i}A_{i} &;& A_{i}B_{i}
\xrightarrow{{p_{2}}} A_{i}E_{i}, \label{spr3} \\ 
A_{i} \xrightarrow{{d_{1}}} E_{i} &;& B_{i} \xrightarrow{{d_{2}}} E_{i}\,.
\label{spr5}
\end{eqnarray}
All constituents have a subscript $i$ to denote that they are located in patch 
$i$. Eq.~(\ref{spr1}) describes the birth of a prey individual, which occurs 
at a rate $b$. We assume that ``space'' is required for this to occur. Also
we do not specify the birth of predator individuals as a separate event, since 
these also occur through predation, as described by Eq.~(\ref{spr3}), and will 
not lead to new terms in the evolution equations. Two types of predation are 
required in Eq.~(\ref{spr3}) so that only a fraction of the resources 
obtained from consumption of the prey are used to produce new predator 
individuals. Finally, Eq.~(\ref{spr5}) describes the death of individuals 
of species $A$ and $B$ at rates $d_1$ and $d_2$ respectively.

Here we are considering an explicitly spatial model, so the additional 
feature which we include is the possibility of changes in the populations 
due to migrations between nearest neighbor patches. These events can be 
described  by adding the following set of reactions\,\cite{AJMaTJN04}:
\begin{eqnarray}
A_{i}E_{j} \xrightarrow{{\mu_{1}}} E_{i}A_{j} &;&
B_{i}E_{j} \xrightarrow{{\mu_{2}}} E_{i}B_{j}\,, \nonumber \\
A_{j}E_{i} \xrightarrow{{\mu_{1}}} E_{j}A_{i} &;& 
B_{j}E_{i} \xrightarrow{{\mu_{2}}} E_{j}B_{i}\,.
\label{spm}
\end{eqnarray}
Here $i$ and $j$ are nearest neighbor sites and $\mu_1$ and $\mu_2$ are the
migration rates for individuals of species $A$ and $B$ respectively.

The state of the system at any given time is specified by the elements of the 
set $\left\{ n_{i}, m_{i} : i=1,\ldots,\Omega \right\}$. If we take the 
transition rates between these states to only depend on the current state of
the system, the process will be Markov and can be described by a master 
equation in continuous time. The natural way to define such transition rates
is according to a mass action law: the probability that two constituents
meet is proportional to their current proportions in their respective patches.
The allowed transitions and the rates at which they take place are given by
Eqs.~(\ref{spr1})-(\ref{spm}). Denoting the transition rates from a state 
with $n_{l}$ predators and $m_{k}$ prey to a state with $n'_{l}$ predators 
and $m'_{k}$ prey  by $T_{n'_{l}, m'_{k}|n_{l}, m_{k}}$, then the transition 
rates corresponding to the purely local reactions (\ref{spr1})-(\ref{spr5}) 
are:
\begin{eqnarray} 
T_{n_{i}+1,m_{i}-1|n_{i},m_{i}} & = & p_{1} \frac{2n_{i}m_{i}}{\Omega N}\,, 
\nonumber \\
T_{n_{i},m_{i}+1|n_{i},m_{i}} & = & b \frac{2m_{i}\left(N-n_{i}-m_{i}\right)}
{\Omega N}\,, \nonumber\\
T_{n_{i}-1,m_{i}|n_{i},m_{i}} & = & d_{1} \frac{n_{i}}{\Omega}\,, \nonumber \\
T_{n_{i},m_{i}-1|n_{i},m_{i}} & = & p_{2} \frac{2n_{i}m_{i}}{\Omega N}
+d_{2} \frac{m_{i}}{\Omega}\,.
\label{pr}
\end{eqnarray} 
These are exactly as in the non-spatial form of the model\,\cite{AJMaTJN05},
but with the state variables all having a subscript $i$ to denote these are
the reactions in patch $i$ and an extra factor of $\Omega$ in the denominator 
since there is a choice between any one of the $\Omega$ patches when 
determining the probability of a transition taking place. To lighten the 
notation we have shown the dependence of $T$ only on the subset of variables 
liable to change (in this case those on the site $i$). The corresponding
expressions for the transition rates between nearest neighbors, which 
describes the migratory process, are
\begin{eqnarray}
T_{n_{i}+1,n_{j}-1|n_{i},n_{j}} & = &
\mu_{1}\frac{n_{j}\left(N-n_{i}-m_{i}\right)}{z\Omega N},\nonumber \\
T_{n_{i}-1,n_{j}+1|n_{i},n_{j}} & = &
\mu_{1}\frac{n_{i}\left(N-n_{j}-m_{j}\right)}{z\Omega N},\nonumber \\
T_{m_{i}+1,m_{j}-1|m_{i},m_{j}} & = &
\mu_{2}\frac{m_{j}\left(N-n_{i}-m_{i}\right)}{z\Omega N},\nonumber \\
T_{m_{i}-1,m_{j}+1|m_{j},m_{j}} & = &
\mu_{2}\frac{m_{i}\left(N-n_{j}-m_{j}\right)}{z\Omega N}\,.
\label{nn}
\end{eqnarray} 
Here, $z$ denotes the coordination number of the lattice, that is the number of
nearest neighbors of any given site, which in our case is $2d$. It needs to be
included since it represents the choice of nearest neighbor $j$, once the 
patch $i$ has been chosen. 

The master equation which governs the time evolution of the system can now be
constructed. Although this equation can easily be written down, and has the
standard form of a sum of transition probabilities giving rise to a change in
the probability distribution function with time\,\cite{NGvK92}, it has a 
rather ungainly appearance. It can be made to look neater through the 
introduction of a little more notation. First, the probability distribution 
function that the system is in state 
$\left\{ n_{i}, m_{i} : i=1,\ldots,\Omega \right\}$ at time $t$ is 
conventionally denoted by
$P\left(n_{1},m_{1},\dots,n_{\Omega},m_{\Omega};t\right)$, but we will 
denote it by $P_{{\bf n},{\bf m}}(t)$. Then the master equation takes the form
\begin{equation}
\frac{dP_{{\bf n},{\bf m}}(t)}{dt} = \sum_{i=1}^{\Omega}
{\cal T}^{\rm loc}_{i}\,P_{{\bf n},{\bf m}} (t)
+ \sum_{i=1}^{\Omega} \sum_{j \in i} 
{\cal T}^{\rm mig}_{ij}\,P_{{\bf n},{\bf m}} (t)\,,
\label{master}
\end{equation}
where the notation $j \in i$ means that $j$ is a nearest neighbor of $i$ and
where ${\cal T}^{\rm loc}_{i}$ and ${\cal T}^{\rm mig}_{ij}$ are transition
rates which are defined below. These transition rates may in turn be 
simplified by the introduction of the step operators\,\cite{NGvK92}
$E_{x_{i}}^{\pm1}$ and $E_{y_{i}}^{\pm1}$ defined by their effect on a typical
function of ${\bf n}$ and ${\bf m}$ as follows:
\begin{eqnarray}
E_{x_{i}}^{\pm1}f\left(n_{i},m_{i}\right) &=& f\left(n_{i}\pm1,m_{i}\right)\,,
\nonumber \\
E_{y_{i}}^{\pm1}f\left(n_{i},m_{i}\right) &=& f\left(n_{i},m_{i}\pm1\right)\,.
\label{step}
\end{eqnarray}
The local transition operator ${\cal T}^{\rm loc}_{i}$ may now be written as
\begin{eqnarray}
{\cal T}^{\rm loc}_{i} &=& \left(E_{x_{i}}-1\right)
T_{n_{i}-1,m_{i}|n_{i,}m_{i}}\nonumber \\ 
&+& \left(E_{y_{i}}^{-1}-1\right)T_{n_{i},m_{i}+1|n_{i},m_{i}} \nonumber \\
&+& \left(E_{y_{i}}-1\right)T_{n_{i},m_{i}-1|n_{i},m_{i}}\nonumber \\
&+& \left(E_{x_{i}}^{-1}E_{y_{i}}-1\right)T_{n_{i}+1,m_{i}-1|n_{i},m_{i}}\,,
\label{local_contrib} 
\end{eqnarray}
with the four local transition rates given explicitly in Eq.~(\ref{nn}).
Similarly, the transition operator ${\cal T}^{\rm mig}_{ij}$ which involves
transitions between nearest neighbor sites can be written as
\begin{eqnarray}
{\cal T}^{\rm mig}_{ij} &=& \left(E_{x_{i}}^{-1}E_{x_{j}}-1\right)
T_{n_{i}+1,n_{j}-1|n_{i},n_{j}} \nonumber\\ 
&+& \left(E_{x_{i}}E_{x_{j}}^{-1}-1\right)
T_{n_{i}-1,n_{j}+1|n_{i},n_{j}}\nonumber\\
&+& \left(E_{y_{i}}^{-1}E_{y_{j}}-1\right)
T_{m_{i}+1,m_{j}-1|m_{i},m_{j}}\nonumber\\
&+& \left(E_{y_{i}}E_{y_{j}}^{-1}-1\right)
T_{m_{i}-1,m_{j}+1|m_{i},m_{j}}\,.
\label{mig_contrib}
\end{eqnarray}

The master equation (\ref{master}), together with the definitions of the 
transitions rates given by Eqs.~(\ref{pr}) and (\ref{nn}) together with
Eqs.~(\ref{local_contrib}) and (\ref{mig_contrib}), completely define the
model once initial and boundary conditions are specified. The model is far too
complicated to be solved exactly, but it can be analyzed very accurately by
studying it in the limit of large system size. As previously 
proposed\,\cite{AJMaTJN04,AJMaTJN05,alo07,mck07}, and as discussed in 
Appendix A, the leading order in a system-size expansion of the master equation
gives deterministic equations whose stationary state can be analyzed, whereas 
the next-to-leading order result gives linear stochastic differential 
equations, which can be Fourier analyzed. From this we can investigate the 
possible existence of resonant behavior induced by the demographic 
stochasticity of the original model. 

In the next section we analyze the equations describing the model to leading 
order and next-to-leading order in the system-size expansion. The details of 
the calculation required to determine these is given in Appendix A.
 
\section{Deterministic limit and fluctuations about it}
\label{deter}

The deterministic limit of the model defined by Eqs.~(\ref{master}), 
(\ref{local_contrib}) and (\ref{mig_contrib}) is derived in Appendix A. It is
defined in terms of the populations $\phi_{i} = \lim_{N \to \infty} (n_i/N)$ 
and $\psi_{i} = \lim_{N \to \infty} (m_i/N)$ and explicitly given by 
Eqs. (\ref{local_macro_1}), (\ref{local_macro_2}), (\ref{CrossMacro1}) and
(\ref{CrossMacro2}). These may be written as the $2\Omega$ macroscopic 
equations
\begin{eqnarray}
\lefteqn{\frac{d\phi_{i}}{d\tau}=2p_{1}\phi_{i}\psi_{i}-d_{1}\phi_{i}}
\nonumber\\ &
&+\mu_{1} \left(\Delta\phi_{i}+\phi_{i}\Delta\psi_{i}
-\psi_{i}\Delta\phi_{i}\right)\,,
\label{dl1}\\
\lefteqn{\frac{d\psi_{i}}{d\tau}=-2\left(p_{1}+p_{2}+b\right)\phi_{i}\psi_{i}}
\nonumber\\
&
&+\left(2b-d_{2}\right) \psi_{i} - 2b \psi_{i}^{2}\nonumber\\
&
&+\mu_{2} \left(\Delta\psi_{i}+\psi_{i}\Delta\phi_{i}
-\phi_{i}\Delta\psi_{i}\right)\,,
\label{dl2}
\end{eqnarray}
where $i=1,\ldots,\Omega$ and where the symbol $\Delta$ represents the 
discrete Laplacian operator 
$\Delta f_i=\frac{2}{z}\sum_{j\in i}\left(f_j-f_i\right)$. A rescaled time,
$\tau = t/\Omega$, has also been introduced.

To complete the formulation of the problem, initial and boundary data should 
be provided. For the type of system considered here the most natural choice 
is to consider zero-flux boundary conditions, regardless of the initial 
conditions. This corresponds to the condition that individuals are not 
allowed to leave or enter the fixed region designated as the system, in other
words there is no immigration or emigration. The system of equations 
(\ref{dl1})-(\ref{dl2}) possesses two limits of interest. The limit $\Omega=1$ 
formally corresponds to a one-site system and is simply the well-known 
Volterra model as studied in \cite{AJMaTJN05}. The limit $\Omega \to \infty$
corresponds to shrinking the lattice spacing to zero and so obtaining a 
continuum description in which the discrete Laplacian operator is replaced
by the continuous Laplacian $\nabla^2$ and the Eqs.~(\ref{dl1})-(\ref{dl2})
become a pair of partial differential equations:
\begin{eqnarray}
\label{pde1}
\frac{\partial\phi}{\partial\tau} & = &
\alpha\phi\psi-\beta\phi+\mu_{1}\nabla^{2}\phi \nonumber\\ 
& &+\mu_{1}\left(\phi\nabla^{2}\psi-\psi\nabla^{2}\phi\right)\,, \\
\frac{\partial\psi}{\partial\tau} & = &
r\psi\left(1-\frac{\psi}{K}\right)-\lambda\psi\phi+
\mu_{2}\nabla^{2}\psi \nonumber \\
& &+\mu_{2}\left(\psi\nabla^{2}\phi-\phi\nabla^{2}\psi\right)\,,
\label{pde2}
\end{eqnarray}
where $\alpha=2p_{1}$, $\beta=d_{1}$, $r=2b-d_{2}$,
$K=(2b-d_2)/2b$, and $\lambda=2\left(p_{1}+p_{2}+b\right)$,
with $\psi$ and $\phi$ representing the prey and predators densities
respectively. It should be noted that in the transition to a continuum model,
the population fractions go over to population densities and parameters may
be scaled by factors involving the lattice spacing. An example of this 
involves the migration rates in Eqs.~(\ref{pde1})-(\ref{pde2}), which are 
scaled versions of those appearing in Eqs.~(\ref{dl1})-(\ref{dl2}) 
(see Eqs.~(\ref{scaled_mig_rates})).

One of the most interesting features of Eqs.~(\ref{pde1})-(\ref{pde2}) is the 
emergence of cross-diffusive terms of the type 
$(\psi\nabla^{2}\phi-\phi\nabla^{2}\psi)$. These types of contributions do not 
usually appear in the heuristically proposed spatially extended predator-prey 
models \cite{AO80b,EEH94}. However, they seem to appear naturally in these 
types of lattice models, and cross-diffusive terms similar to those found 
here have been obtained as the mean-field limit of a set of models proposed 
by Satulovsky \cite{JES96}. An inspection of Eqs.~(\ref{pde1})-(\ref{pde2}) 
leads to the conclusion that they do not reduce to a simple reaction-diffusion 
scheme for any choice of parameters, however if zero-flux boundary conditions 
are chosen, this implies that, after a single integration over the spatial 
domain, the contribution of the cross-diffusive terms for the solution 
vanishes:
\begin{eqnarray}
\int_{A}\left[\phi\nabla^{2}\psi-\psi\nabla^{2}\phi\right]dA'
&=&\int_{C}\left[\phi\nabla\psi-\psi\nabla\phi\right]\cdot
d{\bf r}\nonumber\\ &=&0\,,
\label{flux_van}
\end{eqnarray}
with a similar equation with $\phi$ and $\psi$ interchanged. The condition 
(\ref{flux_van}) also occurs if we impose the requirement that
$\psi({\bf{r}},t)$ and $\phi({\bf{r}},t)$ vanish as 
${\bf{r}}\rightarrow\infty$, instead of the zero-flux boundary conditions, 
which are those typically chosen in textbooks\,\cite{JDM89}.

Before discussing the equations which describe the stochastic behavior of the
system, we will analyze the nature of the stationary solutions in the
deterministic limit. We will be particularly interested in investigating the 
possibility that ``diffusion-driven'' instabilities may occur for the model
defined by Eqs.~(\ref{dl1})-(\ref{dl2}) or equivalently for 
Eqs.~(\ref{pde1})-(\ref{pde2}).

\subsection{Stationary state in the deterministic limit.}
\label{Turing}
One of the simplest questions one can ask about Eqs.~(\ref{dl1})-(\ref{dl2})
or Eqs.~(\ref{pde1})-(\ref{pde2}) concerns the nature of the stationary state. 
It is simple to verify that there are two unstable fixed points (describing 
the null state, $\phi^{*}=\psi^{*}=0$, and a state with no predators, 
$\phi^{*}=0$, $\psi^{*}=K$), and a single coexistence fixed point given by 
(see also\,\cite{FR76,JJ76,SRD83}, for instance)
\begin{equation}
\phi^{*} = \frac{r}{\lambda}\left(1-\frac{\beta}{\alpha K}\right)\,, 
\ \ \psi^{*} = \frac{\beta}{\alpha}\,.
\label{fixed-points}
\end{equation}

Finding non-homogeneous stationary state solutions would require solving a
pair of coupled non-linear differential equations, but we can look for 
solutions if the homogeneous solutions (\ref{fixed-points}) are unstable to 
spatially inhomogeneous small perturbations. That is, we look for solutions of 
Eqs.~(\ref{dl1})-(\ref{dl2}) which have the form
\begin{equation}
\phi_{j} = \phi^{*} + u_{j}\,, \ \ 
\psi_{j} = \psi^{*} + v_{j}\,,
\label{spatial_pert}
\end{equation}
where $u_j$ and $v_j$ are the small perturbations. An exactly similar analysis 
could be carried out on the continuum versions (\ref{pde1})-(\ref{pde2}), but
now $u$ and $v$ would be functions of ${\bf r}$, a vector in the region of 
interest. Substituting Eq.~(\ref{spatial_pert}) into 
Eqs.~(\ref{dl1})-(\ref{dl2}), and keeping only linear terms in $u$ and $v$ 
gives
\begin{eqnarray}
\label{LSA1}
\frac{du_j}{d\tau} &=& a_{11} u_j + a_{12} v_j + \mu_{1} \Delta u_j 
\nonumber \\
& & +\mu_{1}\left(\phi^{*}\Delta v_j -\psi^{*}\Delta u_j\right)\,, \\
\frac{dv_j}{d\tau} & = &
a_{21} u_j + a_{22} v_j + \mu_{2}\Delta v_j \nonumber \\
& &+\mu_{2}\left(\psi^{*}\Delta u_j -\phi^{*}\Delta v_j\right)\,.
\label{LSA2}
\end{eqnarray}
Here $a_{11}, a_{12}, a_{21}$ and $a_{22}$ are the contributions which would 
be found if the perturbation had been assumed to be homogeneous; they are 
exactly the terms found in \cite{AJMaTJN05}, namely
\begin{eqnarray}
a_{11} = \alpha \psi^{*} - \beta; \ \ a_{12} &=& \alpha \phi^{*}; \nonumber \\
a_{21} = - \lambda\psi^{*}; \ \ a_{22} &=& r\left( 1 - \frac{2\psi^*}{K}\right)
- \lambda \phi^{*}\,.
\label{homo_pert}
\end{eqnarray}

We may write Eqs.~(\ref{LSA1}) and (\ref{LSA2}) in the unified form
$\dot{{\bf{u}}}_j=\mathbb{A}{\bf{u}}_j$ with ${\bf{u}}_j=(u_j,v_j)^T$ for 
a given site $j$. The entries of the matrix $\mathbb{A}$ will be denoted 
by $\alpha_{i,11}, \alpha_{i,12}, \alpha_{i,21}$ and $\alpha_{i,22}$. 
The solution to $\dot{{\bf{u}}}_j=\mathbb{A}{\bf{u}}_j$ has the form
\begin{equation}
{\bf u}_j (\tau) \sim \exp\{ \nu \tau + i a {\bf k}. {\bf j} \}\,,
\label{form_of_soln}
\end{equation}
where $a$ is the lattice spacing and where we have explicitly indicated the
vector nature of ${\bf j}$ and ${\bf k}$. The $\nu$ and ${\bf k}$ must 
satisfy
\begin{equation}
\left|
\begin{array}{cc}
\nu - \alpha_{11} & -\alpha_{12} \\
- \alpha_{21} & \nu - \alpha_{22}
\end{array}
\right| = 0\,,
\label{eigenvalue_cond}
\end{equation}
where
\begin{eqnarray}
\alpha_{{\bf k}, 11} &=& a_{11}+\mu_{1}\left(1-\psi^{*}\right) 
\Delta_{\bf k}\,,\nonumber\\
\alpha_{{\bf k}, 12} &=& a_{12}+\mu_{1}\phi^{*} \Delta_{\bf k}\,, \nonumber \\
\alpha_{{\bf k}, 21} &=& a_{21}+\mu_{2}\psi^{*} \Delta_{\bf k}\,, \nonumber \\
\alpha_{{\bf k}, 22} &=& a_{22}+\mu_{2}\left(1-\phi^{*}\right)
\Delta_{\bf k}\,,
\label{alphas}
\end{eqnarray}
and where the discrete Laplacian, $\Delta_{\bf k}$ for a $d-$dimensional 
hypercubic lattice is (see Appendix \ref{VK})
\begin{equation}
\Delta_{\bf k} = \frac{2}{d}\,\sum^{d}_{\gamma = 1} 
\left[ \cos(k_{\gamma} a) - 1 \right]\,.
\label{dis_Lap_k}
\end{equation}

The idea that patterns can form due to a diffusion-induced instability was 
first put forward by Turing in 1952 in connection with his investigation into
the origins of morphogenesis\,\cite{tur52}. More generally, such patterns can
arise in reaction-diffusion equations where a homogeneous stationary state is
stable to homogeneous perturbations, but where irregularities or stochastic 
fluctuations in real systems can induce local deviations from the spatially
uniform state, which can in turn grow if this state is unstable to
inhomogeneous perturbations. Since Turing's seminal work, the phenomenon has
been studied in many types of reaction-diffusion system, including spatial
predator-prey models\,\cite{seg72,AO80b,MP93,DAFB02}. In contrast to these
previous studies, where the reaction-diffusion equations we postulated 
phenomenologically, we have \textit{derived} our equations from a ILM. 
Moreover they differ from the models considered previously because of the
existence of non-linear diffusive terms. Therefore it is of interest to study
if the model we have derived allows for the existence of Turing patterns.

We first need to check that the homogeneous stationary state is stable to
homogeneous perturbations. A homogeneous perturbation means that the $u_j$ and 
$v_j$ in Eq.~(\ref{spatial_pert}) are independent of $j$. This in turn means 
that the terms involving $\mu_1$ and $\mu_2$ are absent from Eqs.~(\ref{LSA1}) 
and (\ref{LSA2}). Therefore the stability to homogeneous perturbations may be
found from Eq.~(\ref{eigenvalue_cond}) with the $\alpha$ replaced by the 
$a$. Stability is assured if $a_{11} + a_{22} < 0$ and 
$a_{11} a_{22} - a_{12} a_{21} > 0$, since these conditions are equivalent to 
asking that the $\nu$ which are solutions of Eq.~(\ref{eigenvalue_cond}) have
negative real parts. It is straightforward to check from the
explicit forms (\ref{fixed-points}) and (\ref{homo_pert}) that $a_{11}=0$,
$a_{12} > 0$ and $a_{21}, a_{22} < 0$, and so that this is the case. As an 
aside we can also check that for the null state ($\phi^{*}=\psi^{*}=0$) and
the state without predators ($\phi^{*}=0$, $\psi^{*}=K$), under the condition
that the fixed point (\ref{fixed-points}) exists, that $a_{11}a_{22}<0$
and $a_{12}=0$. Therefore the determinant of the stability matrix is negative,
and so the eigenvalues are real with different signs, and both these states are
unstable.

To get a diffusive instability, we need to investigate the solutions 
(\ref{spatial_pert}) which now include the spatial contributions. For an 
instability to occur, one of the conditions ${\rm tr}\mathbb{A}_{\bf k} < 0$ 
or ${\rm det}\mathbb{A}_{\bf k} > 0$ must be violated. From 
Eq.~(\ref{dis_Lap_k}) it is clear that $\Delta_{\bf k} \leq 0$ and so from 
Eq.~(\ref{alphas}) that $\alpha_{11} \leq a_{11}$ and $\alpha_{22} \leq a_{22}$
and so that ${\rm tr}\mathbb{A}_{\bf k} < 0$. So the only possibility for a 
Turing pattern to arise is if ${\rm det}\mathbb{A}_{\bf k} < 0$. By direct 
calculation
\begin{eqnarray}
{\rm det}\mathbb{A}_{\bf k} &=& - a_{12}a_{21} - \mu_{1}\left[ a_{21}\phi^{*} 
- a_{22}\left( 1 - \psi^{*} \right) \right]\Delta_{\bf k} \nonumber \\
&-& \mu_{2} a_{12} \psi^{*}\Delta_{\bf k} + \mu_{1}\mu_{2}\left(1-\phi^{*} 
-\psi^{*} \right)\Delta^{2}_{\bf k}\,.
\label{det_A} 
\end{eqnarray}
Now all the terms on the right-hand side of Eq.~(\ref{det_A}) are manifestly
positive, except the second. However, since
\begin{equation}
a_{21}\phi^{*} - a_{22}\left( 1 - \psi^{*} \right) = r\psi^{*} \left( \frac{1}
{K} - 1 \right)\,,
\label{second_term}
\end{equation}
and $K=1 - (d_{2}/2b) < 1$, then this term is also positive. Therefore
${\rm det}\mathbb{A}_{\bf k} > 0$ and so the homogeneous stationary state is 
stable to both small homogeneous and small inhomogeneous perturbations.

It has been known for some time that the simple reaction-diffusion equations 
for a predator-prey model (i.e. those containing only containing simple
diffusive terms such as $\nabla^{2}\phi$ and $\nabla^{2}\psi$) do not lead
to diffusive instabilities\,\cite{JDM89}. We have shown here that the 
introduction of a particular type of cross-diffusive term, which has its 
origins in the ILM formulation, also contains no Turing instability. It should
be noted that this also holds true in the limit of zero lattice spacing where
$\Delta_{\bf k}$ is replaced by $-k^{2}$ (up to a constant), which is also 
always negative for ${\bf k} \neq 0$. This corresponds to using 
Eqs.~(\ref{pde1})-(\ref{pde2}), rather than Eqs.~(\ref{dl1})-(\ref{dl2}). 
Since, on average, the population fractions do not exhibit any form of spatial 
self-organizing structure, the emergence of such structures when observing the 
full dynamical process should be understood as an effect due to fluctuations. 
So we now study the next next-to-leading order contributions which describe 
fluctuations around these mean values, with the aim of quantifying possible 
resonant behavior in both space and time.

\subsection{Fluctuations}
\label{linear_noise}

The next-to-leading order in the system size expansion gives a Fokker-Planck 
equation in the $2\Omega$ variables $\xi_i$ and $\eta_i$, which describe the 
deviation of the system from the mean fields:
\begin{equation}
\xi_i(t) = \sqrt{N} \left( \frac{n_i}{N} - \phi_i(t) \right)\,, \ \ 
\eta_i(t) = \sqrt{N} \left( \frac{m_i}{N} - \psi_i(t) \right)\,.
\label{deviations}
\end{equation}
The equation itself is derived in Appendix \ref{VK}; it is given by 
Eq.~(\ref{FPE}) with coefficients defined by Eqs.~(\ref{A_def}) and 
(\ref{B_def}). These coefficients have been evaluated at the fixed-point
$\phi^*$, $\psi^*$ of the deterministic equations since, as explained earlier,
we are interested in studying the effect of fluctuations on the system once 
transient solutions of the deterministic equations have died away. Rather 
than work with this Fokker-Planck equation, it is more convenient to use the 
Langevin equation which it is equivalent to. This has the 
form\,\cite{ris89,gar04}
\begin{equation}
\frac{d\boldsymbol{\zeta}_i}{d\tau} = {\cal A}_{i}(\boldsymbol{\zeta}) 
+ \boldsymbol{\lambda}_{i}(\tau)\,,
\label{Langevin}
\end{equation}
where
\begin{equation}
\langle \boldsymbol{\lambda}_{i}(\tau) 
\boldsymbol{\lambda}_{j}(\tau') \rangle = {\cal B}_{ij} \delta(\tau-\tau')\,.
\label{correlators}
\end{equation}
Here ${\boldsymbol{\zeta}}_{i}=(\xi_{i}, \eta_{i})$ and 
${\boldsymbol{\lambda}}_{i}=(\lambda_{i,1}, \lambda_{i,2})$ with 
${\cal B}_{ij}$ being the constant matrix defined by Eq.~(\ref{B_def}). 

The key point here is that the system-size expansion to this order yields a
function ${\cal A}(\boldsymbol{\zeta})$ which is \textit{linear} in 
$\boldsymbol{\zeta}_i$, as can be seen from Eq.~(\ref{A_def}). It is this 
linear nature of the Langevin equation which is crucial in the analysis that 
follows. To study possible cyclic behavior we will require to calculate the 
power spectrum of the fluctuations (\ref{deviations}), and to do this we need 
to find an equation for their temporal Fourier transforms. The linearity of 
the Langevin equation (\ref{Langevin}) means that this is readily achieved. 
The translational invariance of the solutions of the deterministic equations, 
together with the nature of the diffusive terms also make it useful to take 
the spatial Fourier transform of Eq.~(\ref{Langevin}). This is discussed in 
detail in Appendix \ref{FT}; writing out the two components of the equation 
explicitly it has the form
\begin{eqnarray}
\frac{d\xi_{\bf k}}{d\tau} &=& \alpha_{{\bf k}, 11}\,\xi_{\bf k} 
+ \alpha_{{\bf k}, 12}\,\eta_{\bf k} + \lambda_{1,{\bf k}}(\tau) \nonumber \\
\frac{d\eta_{\bf k}}{d\tau} &=& \alpha_{{\bf k}, 21}\,\xi_{\bf k} 
+ \alpha_{{\bf k}, 22}\,\eta_{\bf k} + \lambda_{2,{\bf k}}(\tau)\,,
\label{Lang}
\end{eqnarray}
where the $\alpha_{\bf k}$ are given by Eq.~(\ref{alphas}) and by 
Eq.~(\ref{homo_pert}). The noise correlators (\ref{correlators}) are now local
in ${\bf k}-$space:
\begin{equation}
\langle {\bf{\lambda}}_{\bf k}(\tau) {\bf{\lambda}}_{\bf k'}(\tau') \rangle 
= {\cal B}_{\bf k}\,\Omega a^{d}\,\delta_{{\bf k}+{\bf k}', 0} 
\delta(\tau-\tau')\,,
\label{corr}
\end{equation}
where ${\cal B}_{\bf k}$ is derived in the Appendices (see Eq.~(\ref{FT_corr1})
et seq.) and is given by
\begin{eqnarray}
{\cal B}_{{\bf k},11} &=& a^{d}\,\left[ \left(d_{1}\phi^{*} 
+ 2p_{1} \psi^{*}\phi^{*}\right) \right. \nonumber \\
&-& \left. 2 \mu_{1} \phi^{*} \left( 1 - \phi^{*} - \psi^{*} \right)
\Delta_{\bf k} \right]\,, 
\nonumber \\
{\cal B}_{{\bf k},22} &=& a^{d}\,\left[ 2b \psi^{*} \left(1-\phi^{*}
- \psi^{*}\right) + d_{2}\psi^{*} \right. \nonumber \\
&+& \left. 2 \left( p_{1} + p_{2} \right)\psi^{*}\phi^{*} 
- 2 \mu_{2} \psi^{*} \left( 1 - \phi^{*} - \psi^{*} \right)
\Delta_{\bf k} \right]\,, \nonumber \\
{\cal B}_{{\bf k},12} &=& {\cal B}_{{\bf k},21} = -2 a^{d} p_{1} \phi^{*}
\psi^{*}\,.
\label{B_k}
\end{eqnarray}
It should be noted that, since $\Delta_{\bf k} < 0$, the diagonal elements of 
${\cal B}_{\bf k}$ are all positive, as they should be. 

It is interesting to consider what happens in the continuum limit $a \to 0$.
For non-zero $a$, the wave-numbers take on values in the interval 
$(-\pi/a) \leq k_{i} \leq (\pi/a)$, but this becomes an infinite interval as 
$a \to 0$. The wave-numbers are still discrete however, due to the finite 
volume (area in two dimensions) of the system; we keep the volume 
$\Omega a^{d}$ fixed in the limit, so that $\Omega \to \infty$. In the limit 
$\Omega a^{d}\,\delta_{{\bf k}+{\bf k}', 0}$ goes over to
$(2\pi)^{d}\,\delta ({\bf k}+{\bf k}')$ and $\Delta_{\bf k}$ goes over to
$-k^{2}$, as long as the migration rates are suitably scaled 
(see Eq.~(\ref{scaled_mig_rates})). However from Eq.~(\ref{B_k}), it is clear 
that the ${\cal B}_{\bf k}$ vanish in the limit due to the factor of $a^{d}$. 
This should not be too surprising: since $\Omega \to \infty$, the number of 
degrees of freedom of the system is becoming infinitely large, and thus we
would expect fluctuations to vanish. If all the ${\cal B}_{\bf k}$ are zero,
the noises $\boldsymbol{\lambda}_{\bf k}(\tau)$ vanish, and therefore so do
$\xi_{\bf k}(\tau)$ and $\eta_{\bf k}(\tau)$. This effect has been seen
(see \cite{MMITG07} and the references therein): oscillatory behavior in 
these types of models persists as long as the number of sites remains finite, 
however it disappears in the so-called thermodynamic limit. However, in
practice, one has to go over to describing the population sizes as population 
densities, rather than pure numbers, in this limit. This will involve further 
rescalings, and depending on the exact definition of the model, these 
fluctuations can survive the continuum limit. For this reason we will keep a 
finite lattice spacing: the results for a particular continuum model variant 
can then be determined by taking the $a \to 0$ limit in the appropriate manner.

\subsection{Simulations}
\label{sims}

We expect  that the deterministic equations (\ref{dl1}) and (\ref{dl2}), 
together with the stochastic fluctuations about them, given by 
Eqs.~(\ref{Lang})-(\ref{B_k}), will give an excellent description of the
model defined by Eqs.~(\ref{spr1})-(\ref{spm}) for moderate to large system 
size. We test this expectation here by presenting the results of numerical 
simulations performed for the full stochastic process (\ref{spr1})-(\ref{spm})
using the Gillespie algorithm\,\cite{DTG76}. This is completely equivalent to 
solving the full master equation (\ref{master}). To obtain the best results we
restricted our simulations to the one-dimensional system ($d=1$), even though 
our theoretical treatment applies to general $d$ and we would usually be 
interested in $d=2$. We took the length of the spatial interval to be unity,
so that $a\Omega = 1$. Therefore once the number of lattice sites, $\Omega$,
is fixed, so is the lattice spacing, $a$.


\begin{figure*}
\centering \subfigure[]
{\includegraphics[clip,width=1.0\columnwidth,keepaspectratio]{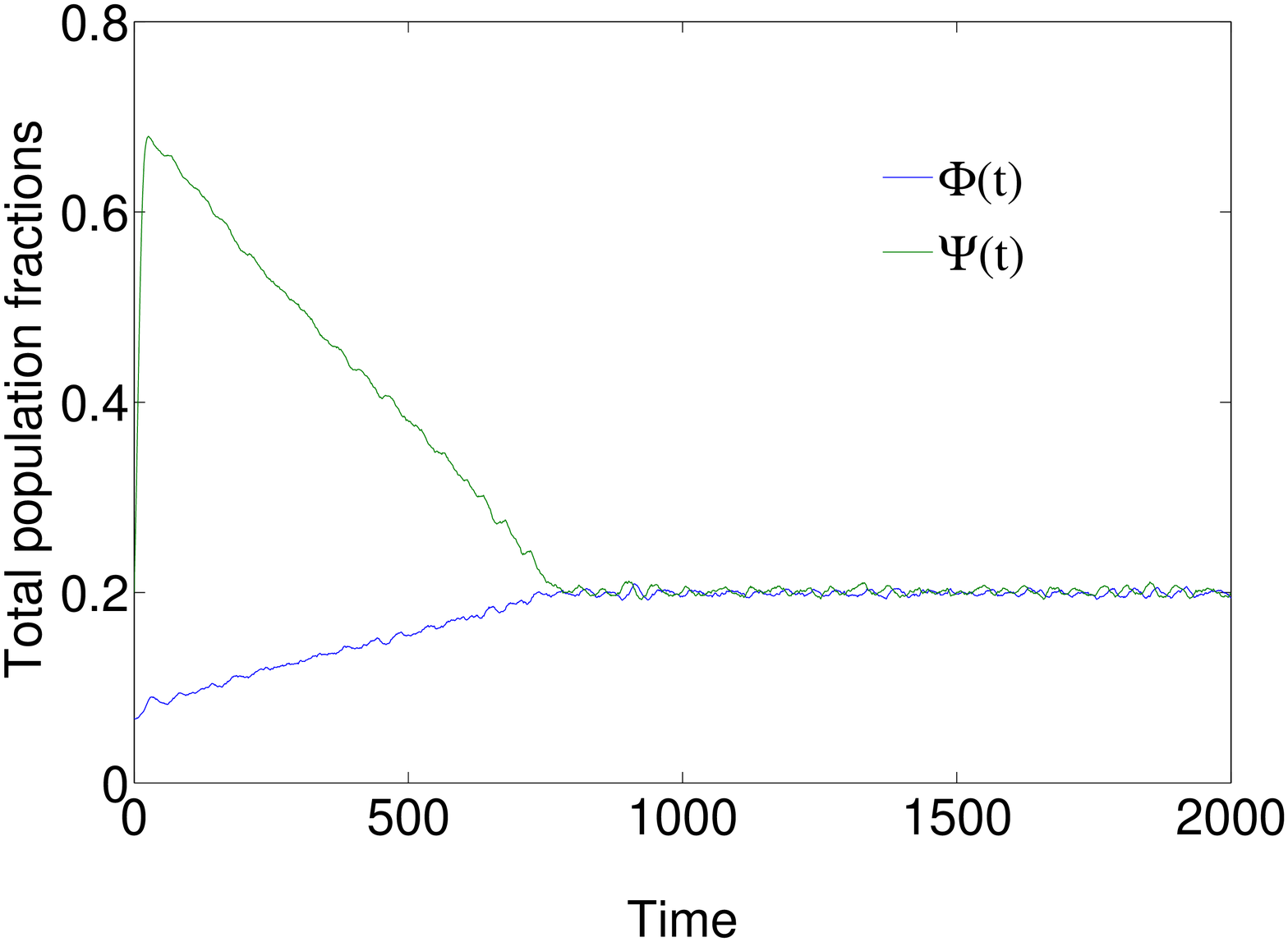}}
\subfigure[]
{\includegraphics[clip,width=1.0\columnwidth,keepaspectratio]{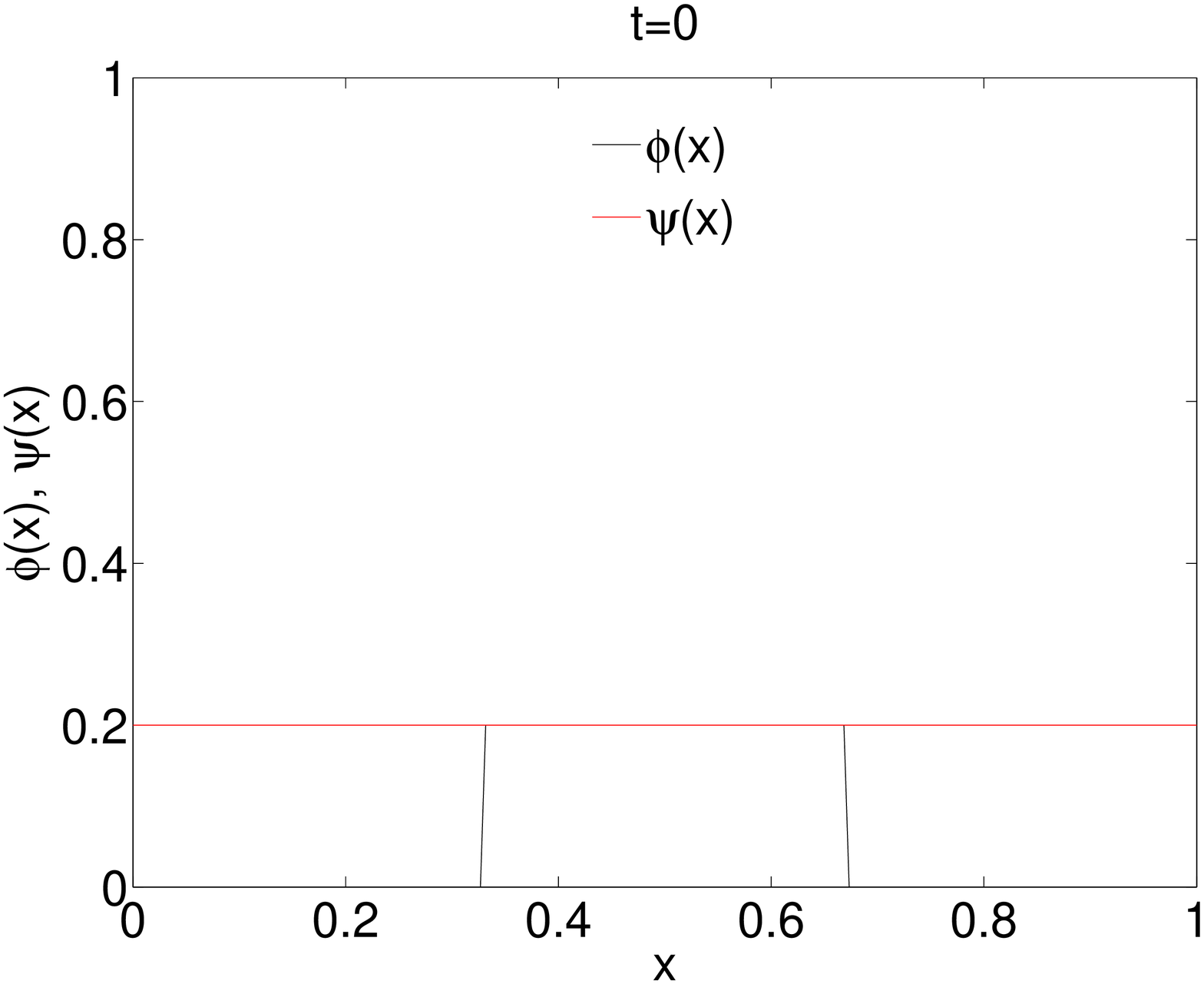}}
\subfigure[]
{\includegraphics[clip,width=1.0\columnwidth,keepaspectratio]{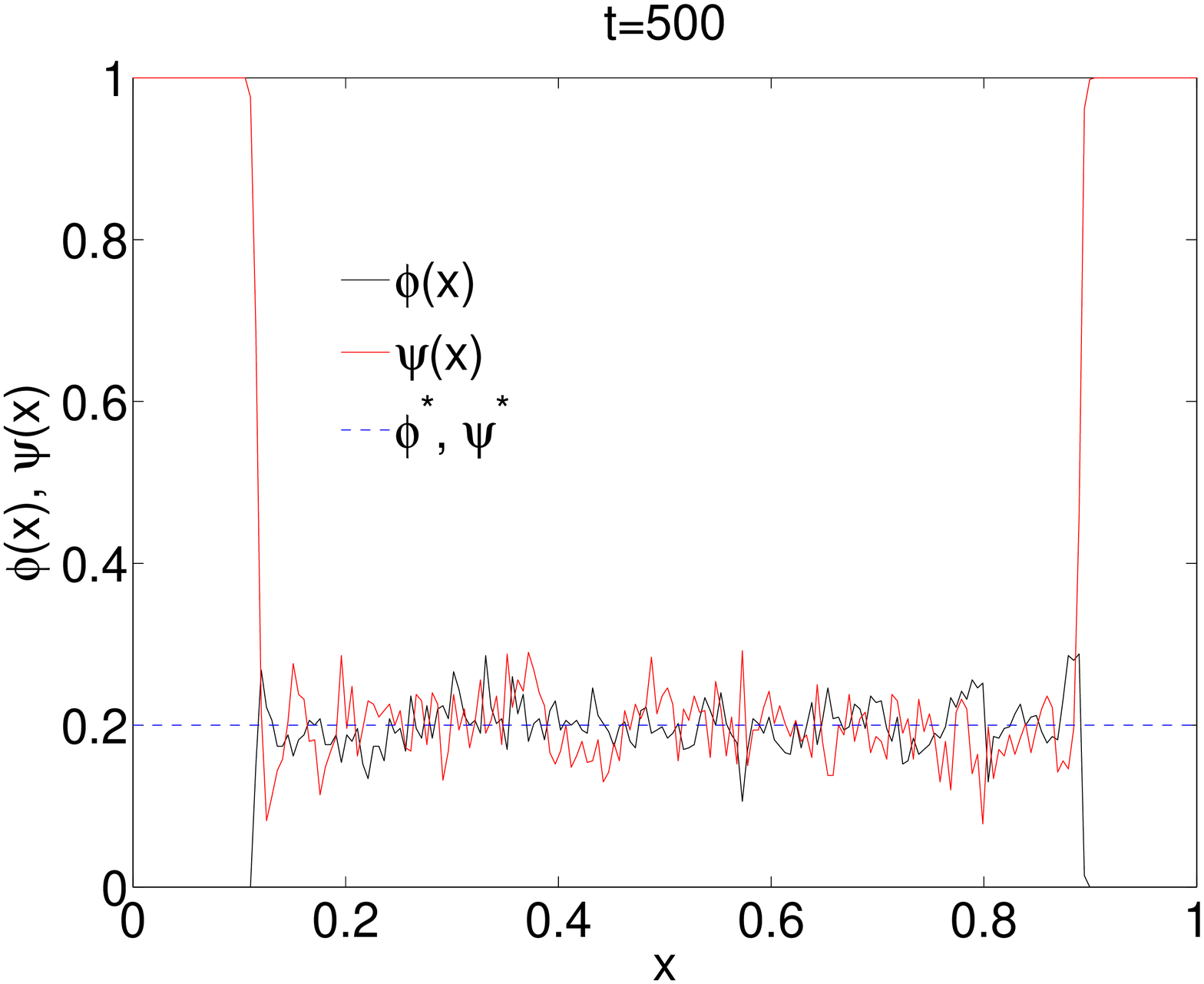}}
\subfigure[]
{\includegraphics[clip,width=1.0\columnwidth,keepaspectratio]{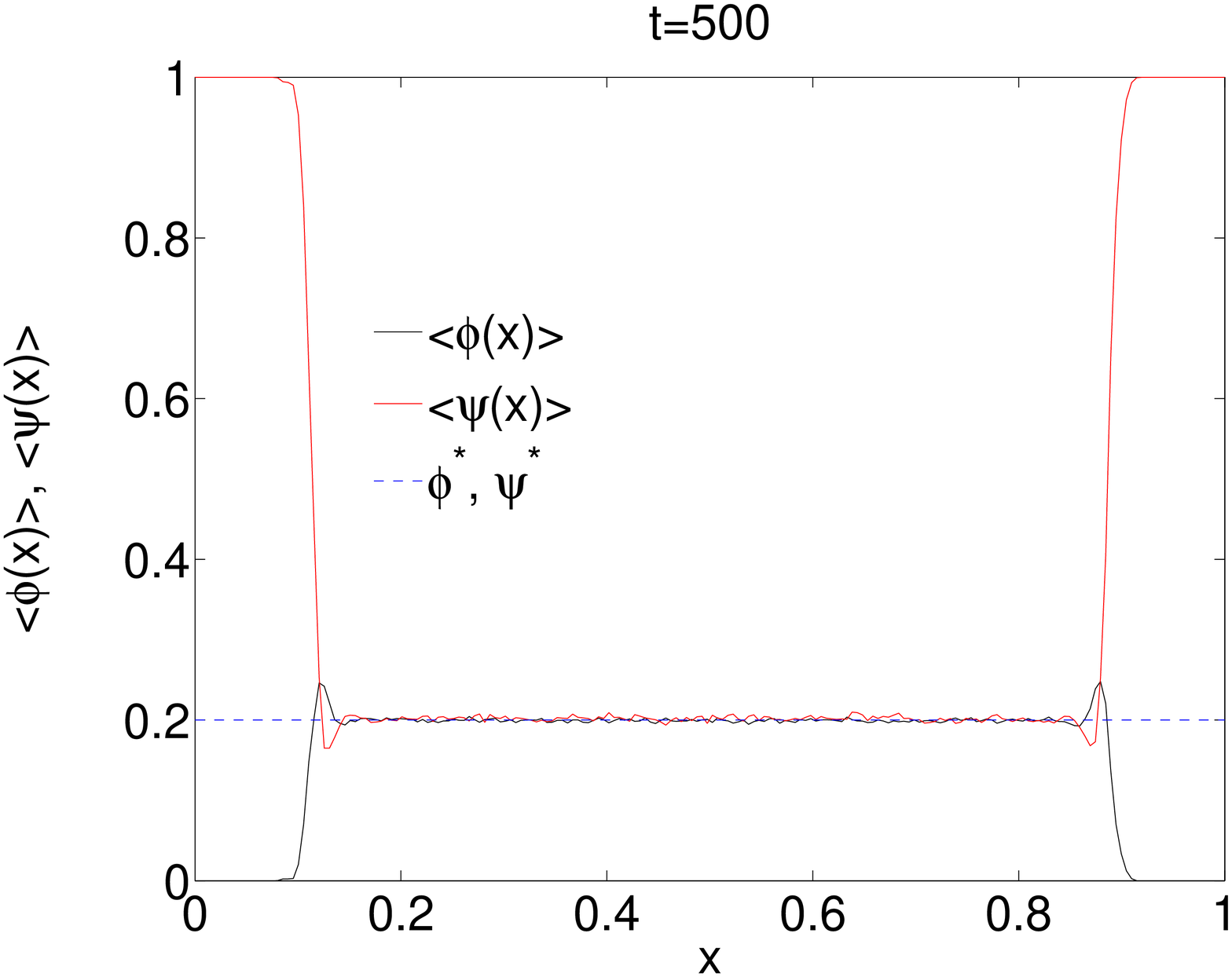}}
\subfigure[]
{\includegraphics[clip,width=1.0\columnwidth,keepaspectratio]{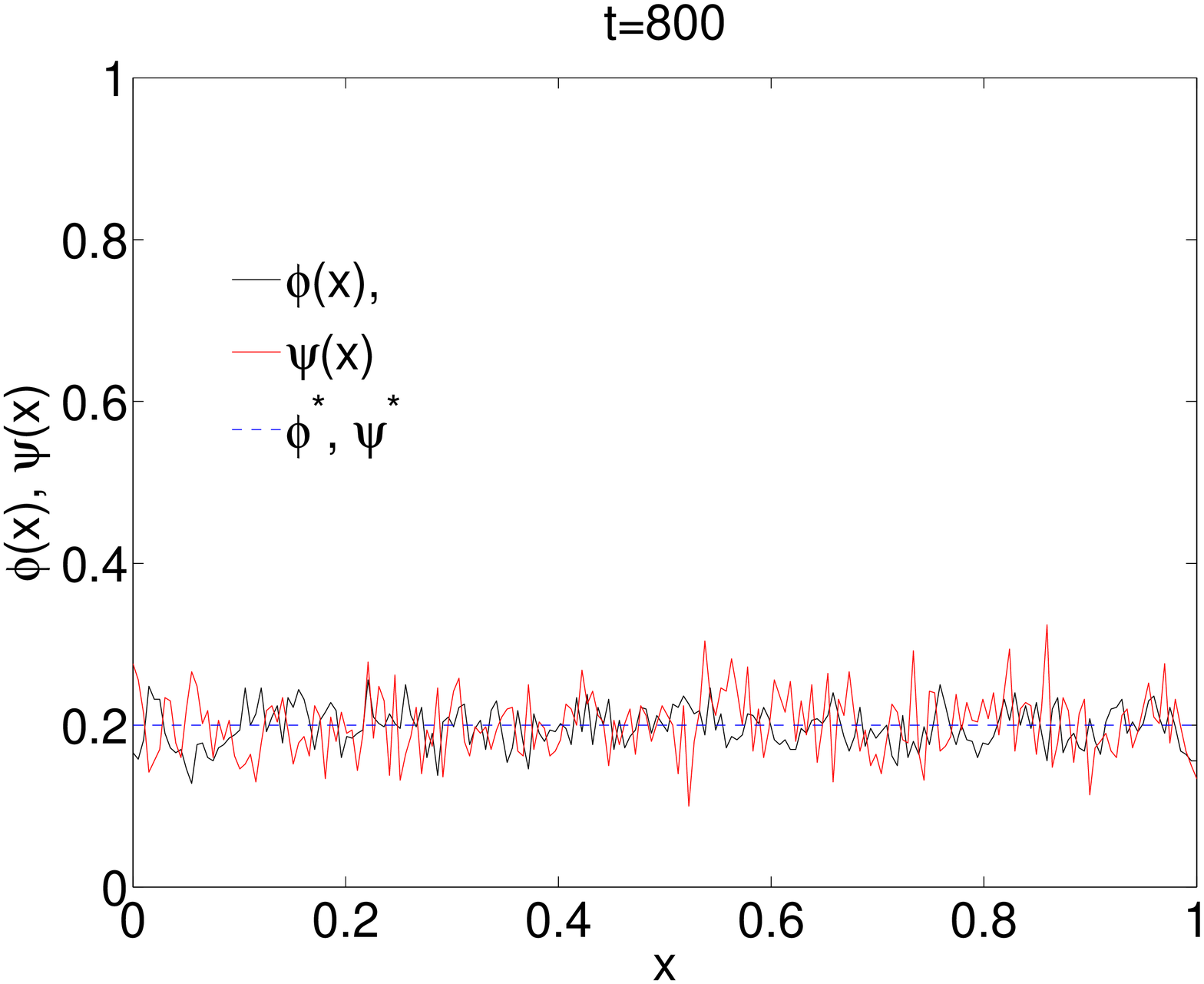}}
\subfigure[]
{\includegraphics[clip,width=1.0\columnwidth,keepaspectratio]{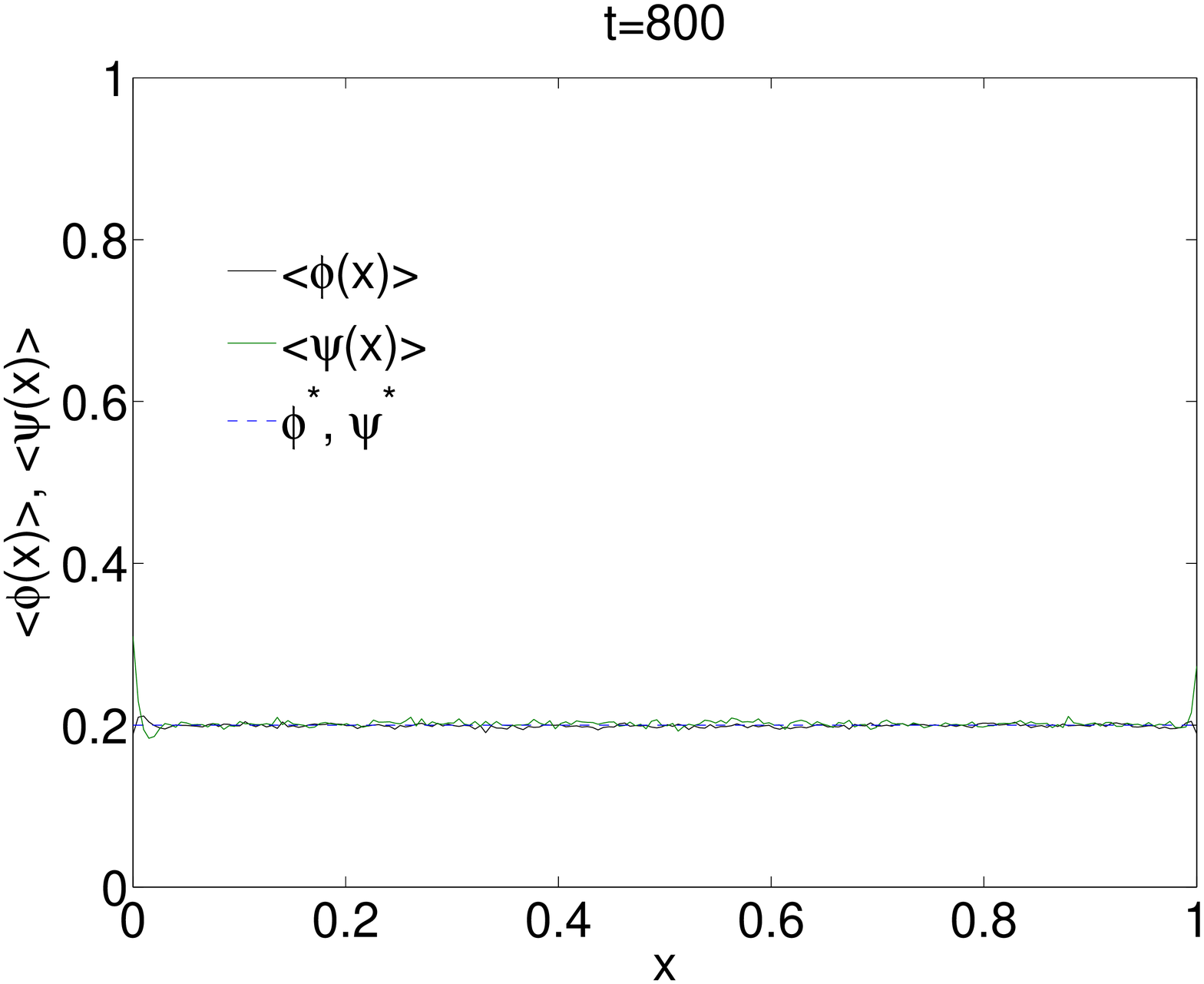}}
\caption{Results obtained from simulating the process
(\ref{spr1})-(\ref{spm}).  Panel (a) shows the temporal evolution of
the total population fractions of predators $\Phi(t)$ and prey
$\Psi(t)$. Panels (b)-(f) contain snapshots at different times of  the
spatial configuration for a typical realization (panels (c) and (e))
and averaging $150$ independent  realizations (panels (d) and (f)). 
Panel (b) shows the initial spatial configuration. The reaction rates
employed were $p_1=0.25\Omega$, $p_2=0.05\Omega$, $d_1=0.1\Omega$, 
$d_2=0.0$, $b=0.1\Omega$, $\mu_1=0.2\Omega$, $\mu_2=0.1\Omega$, 
$\Omega=200$ and $N=500$. The dotted lines in the figure correspond to 
the fixed point values $\phi^*$ and $\psi^*$ found from 
Eq.~(\ref{fixed-points}).}
\label{trun}
\end{figure*}

Panel (a) of Fig.~\ref{trun} shows typical behavior of the total population 
fractions  $\Phi(t)=\frac{1}{\Omega N}\sum_{i=1}^\Omega n_i$ and  
$\Psi(t)=\frac{1}{\Omega N}\sum_{i=1}^\Omega m_i$ starting from the initial 
condition shown in Fig.~\ref{trun}(b). Subsequent panels show the time 
evolution of the local fractions $\phi_i$ and $\psi_i$ starting from the same
initial condition. The time $t$ corresponds to the Gillespie time and was 
measured in integer time-steps. The average values are those calculated
from the fixed point (\ref{fixed-points}). For this simulation the number of 
sites employed was  $\Omega=200$ and the site capacity was $N=500$. The
local reaction rates were chosen so as to match the values used in the
non-spatial version of the  model\,\cite{AJMaTJN05}. In particular this means 
that $\phi^{*}=\psi^{*}$.  Since the time in this spatial version is scaled 
by $\Omega$ ($\tau=t/\Omega$), the rates are $\Omega$ times those used in 
\cite{AJMaTJN05}, namely $p_1=0.25\Omega$, $p_2=0.05\Omega$, $d_1=0.1\Omega$, 
$d_2=0.0$ and $b=0.1\Omega$. The values of the migration rates $\mu_1$ and 
$\mu_2$ for this simulation were $0.2\Omega$ and $0.1\Omega$ respectively.

The initial configuration shown in Fig.~\ref{trun}(b) consists of prey 
homogeneously distributed along the spatial interval, with populations equal 
to the equilibrium coexistence value $m_i=\psi^*N$. The predator species were 
also initially homogeneously distributed, with the difference that they were 
confined to only the middle third of the sites; the first and last third of 
the interval contained no predator individuals. This choice was made in order 
to clearly indicate the nature of the invasion process of predators into the 
predator-free zones, which eventually leads to the establishment of a mixed 
predator-prey regime over the whole spatial interval. Before this happens, 
all those sites with only prey individuals should converge to the saturation 
value $\psi=K$ and remain there until a predator invades the site, which can 
only occur via a migration event. Once the entire spatial interval is 
populated with individuals of the two species, their numbers will oscillate 
around the fixed point $(\phi^*, \psi^*)$, as shown in Figs.~\ref{trun}(c)
and (e). It was found that for the parameter values taken in this realization 
of the process, the mixed state first becomes established in the entire domain 
at approximately $t\sim800$. For times larger than this there is oscillatory 
behavior around the fixed point values, which is shown in later figures
(Figs.~\ref{pwks}(a) and (b)); this behavior resembles that reported in
\cite{AJMaTJN05}.

Figure \ref{trun} also contains a sequence which shows the dynamics of the 
average values (panels (d) and (f)), obtained by averaging over $150$  
independent realizations of the stochastic process. The dynamics consists of a
continuous transition from the unstable state with only prey present, into the 
stable two-species fixed point. This takes the form of traveling wave-fronts 
of ``pursuit'' and ``evasion'',  which describe the invasion process of 
predators into locations occupied only by prey individuals. Such traveling 
waves may be found directly as solutions of the deterministic 
equations \cite{SRD83,JDM89}.

Our main interest in this paper is the study of the nature of the fluctuations
about the stationary state, that is, at times subsequent to that illustrated
in Fig.~\ref{trun}(d), and we now return to their study. 

\section{Power Spectra}
\label{power}
To calculate the power spectra of the fluctuations about the stationary state,
we first have to take the temporal Fourier transform of Eqs.~(\ref{Lang}). This
reduces the equations governing the stochastic behavior of the system to two 
coupled algebraic equations which are linear, and so which can be used to 
obtain a closed form expression for the power spectra. In this section we 
first describe this analytic approach, and then go on to discuss how the power 
spectra can be found from numerical simulations, and then finally compared the 
results of these two approaches.  

\subsection{Analytic form}
\label{power_anal}
Taking the temporal Fourier transform of Eqs.~(\ref{Lang}) yields
\begin{equation}
\mathbb{M}\boldsymbol{\zeta}_{\bf k}(\omega) = \boldsymbol{\lambda}_{\bf k}
(\omega)\,,
\label{double_FT}
\end{equation}
where $\mathbb{M=}\left(-i\omega\mathbb{I}-\mathbb{A}\right)$ and $\mathbb{I}$
is the $2 \times 2$ identity matrix. Therefore 
$\boldsymbol{\zeta} = \mathbb{M}^{-1} \boldsymbol{\lambda}$, which implies 
that
\begin{eqnarray}
\left| \xi_{\bf k}(\omega) \right|^{2} &=& \left| p_{11} \right|^{2} 
\lambda_{1} \lambda^{*}_{1} + p_{11} p_{12}^{*} \lambda_{1} \lambda_{2}^{*}
\nonumber \\
&+& p_{11}^{*} p_{12} \lambda_{1}^{*} \lambda_{2} + \left| p_{22} \right|^{2}
\lambda_{2} \lambda_{2}^{*}\,,
\label{mod_squared}
\end{eqnarray}
with a similar expression for $\left| \eta_{\bf k}(\omega) \right|^{2}$ which
is just Eq.~(\ref{mod_squared}) but with all the first subscripts of $p$ 
changed to $2$. Here the $p_{ab}$ are the elements of $\mathbb{M}^{-1}$. Using
\begin{equation}
\langle \boldsymbol{\lambda}_{\bf k}(\omega) 
\boldsymbol{\lambda}^{*}_{\bf k}(\omega) \rangle = {\cal B}_{\bf k}\,, 
\label{double_corr}
\end{equation}
the power spectra for the predators 
\begin{equation}
P_{{\bf k}, 1} \left( \omega \right) = \left\langle \left| \xi_{\bf k} 
(\omega) \right|^{2} \right\rangle\,,
\label{pred_PS}
\end{equation}
and for the prey
\begin{equation}
P_{{\bf k}, 2} \left( \omega \right) = \left\langle \left| \eta_{\bf k} 
(\omega) \right|^{2} \right\rangle\,,
\label{prey_PS}
\end{equation}
may easily be found.

Since the Langevin equations are diagonal in ${\bf k}-$space, the structure 
of the expressions for the power spectra are the same as those found in other 
studies\,\cite{AJMaTJN05,alo07,mck07}, namely
\begin{equation}
P_{{\bf k}, 1} = \frac{C_{{\bf k}, 1} + {\cal B}_{{\bf k}, 11}\omega^{2}}
{\left[ \left(\omega^{2} - \Omega^{2}_{{\bf k}, 0}\right)^{2} + 
\Gamma_{\bf k}^{2}\omega^{2}\right]}\,,
\label{Pwk1}
\end{equation}
and 
\begin{equation}
P_{{\bf k}, 2} = \frac{C_{{\bf k}, 2} + {\cal B}_{{\bf k}, 22}\omega^{2}}
{\left[ \left(\omega^{2} - \Omega^{2}_{{\bf k}, 0}\right)^{2} + 
\Gamma_{\bf k}^{2}\omega^{2}\right]}\,,
\label{Pwk2}
\end{equation}
where
\begin{eqnarray}
C_{{\bf k}, 1} &=& {\cal B}_{{\bf k}, 11}\,\alpha^{2}_{{\bf k}, 22}
- 2 {\cal B}_{{\bf k}, 12}\,\alpha_{{\bf k}, 12}\,\alpha_{{\bf k}, 22}
+ {\cal B}_{{\bf k}, 22}\,\alpha^{2}_{{\bf k}, 12}\,, \nonumber \\
C_{{\bf k}, 2} &=& {\cal B}_{{\bf k}, 22}\,\alpha^{2}_{{\bf k}, 11}
- 2 {\cal B}_{{\bf k}, 12}\,\alpha_{{\bf k}, 21}\,\alpha_{{\bf k}, 11}
+ {\cal B}_{{\bf k}, 11}\,\alpha^{2}_{{\bf k}, 21}\,. \nonumber \\
\label{Cs}
\end{eqnarray}
The spectra (\ref{Pwk1}) and (\ref{Pwk2}) resemble those found when analyzing 
driven damped linear oscillators in physical systems. A difference between
that situation and the one here is that the driving forces here are white 
noises $\boldsymbol{\lambda}(\tau)$ which excite all frequencies equally, 
thus there is no need to tune the frequency of the ``driving force'' to 
achieve resonance. The parameters in the denominators of Eqs.~(\ref{Pwk1}) and 
(\ref{Pwk2}) are given by $\Omega_{{\bf k}, 0}^2={\rm det}\mathbb{A}_{\bf k}$
and $\Gamma_{\bf k}=-{\rm tr}\mathbb{A}_{\bf k}$, where $\mathbb{A}_{\bf k}$
is the stability matrix found from perturbations about the homogeneous state 
and which has entries given by Eq.~(\ref{alphas}).

We are particularly interested in the situation where there is resonant 
behavior, that is, when there exist particular frequencies when the 
denominators of Eqs.~(\ref{Pwk1}) and (\ref{Pwk2}) are small. The denominator 
vanishes when $(i\omega)^{2}+(i\omega)\,{\rm tr}\mathbb{A}_{\bf k}+
{\rm det}\mathbb{A}_{\bf k}=0$, which never occurs at real values of $\omega$,
however it does occur for complex $\omega$ with non-zero real part if
$({\rm tr}\mathbb{A}_{\bf k})^{2}< 4\,{\rm det}\mathbb{A}_{\bf k}$. This pole
in the complex $\omega-$plane indicates the existence of a resonance, and
is exactly the same condition that the stability matrix $\mathbb{A}_{\bf k}$ 
has complex eigenvalues. This conforms with our intuition that the approach
to the homogeneous stationary state needs to be oscillatory for 
demographic stochasticity to be able to turn this into cyclic behavior. If
the $\omega$ dependence of the spectra numerators is ignored, then it is
simple to show that the spectra have a maximum in $\omega$ if additionally
$({\rm tr}\mathbb{A}_{\bf k})^{2}< 2\,{\rm det}\mathbb{A}_{\bf k}$. Using the
full numerator results in a condition which is only slightly more 
complicated\,\cite{alo07,mck07}.

\subsection{Numerical results}
\label{numer}


\begin{figure*}
\centering  \subfigure[]
{\includegraphics[clip,width=1.0\columnwidth,keepaspectratio]{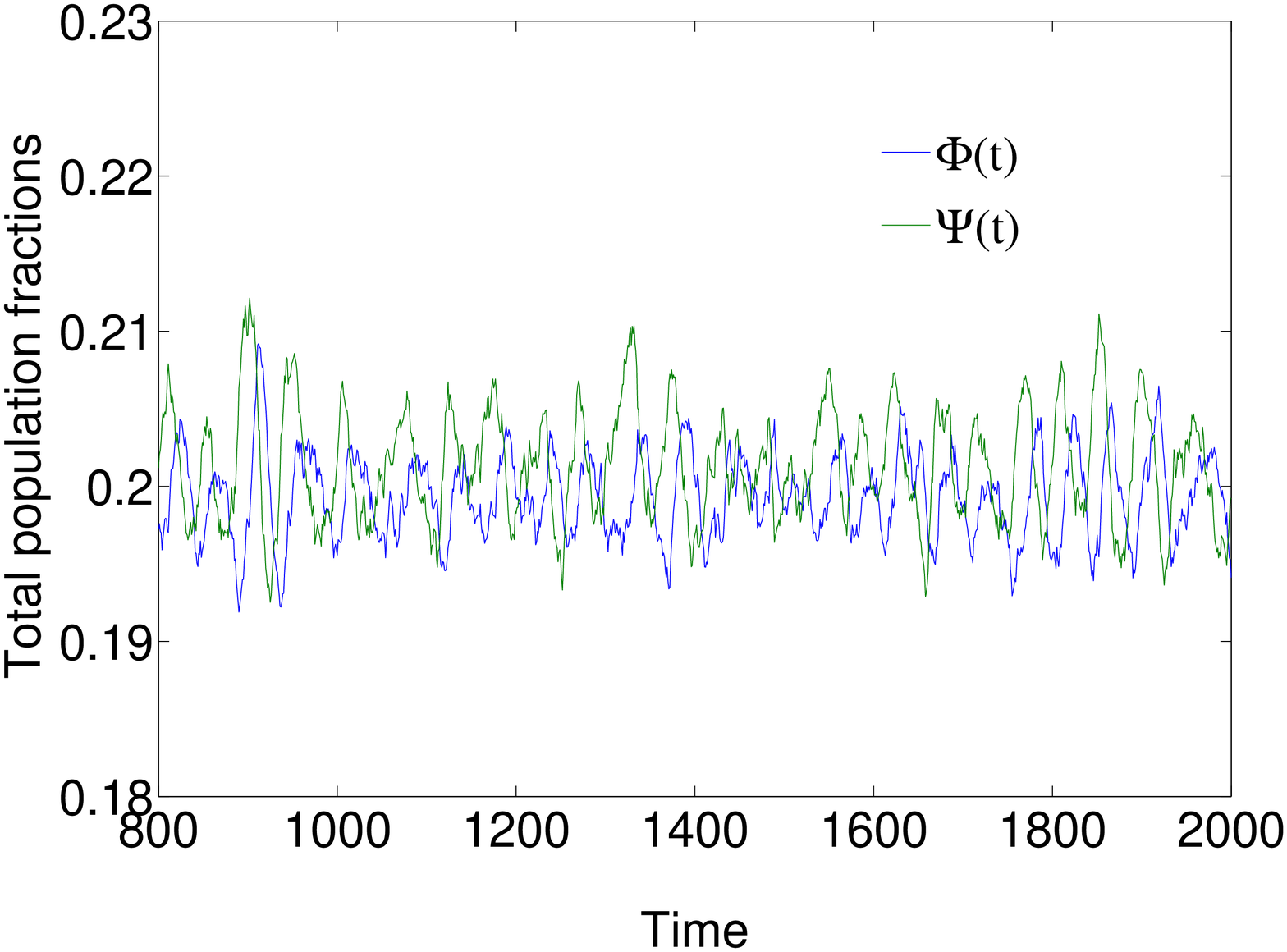}}
\subfigure[]
{\includegraphics[clip,width=1.0\columnwidth,keepaspectratio]{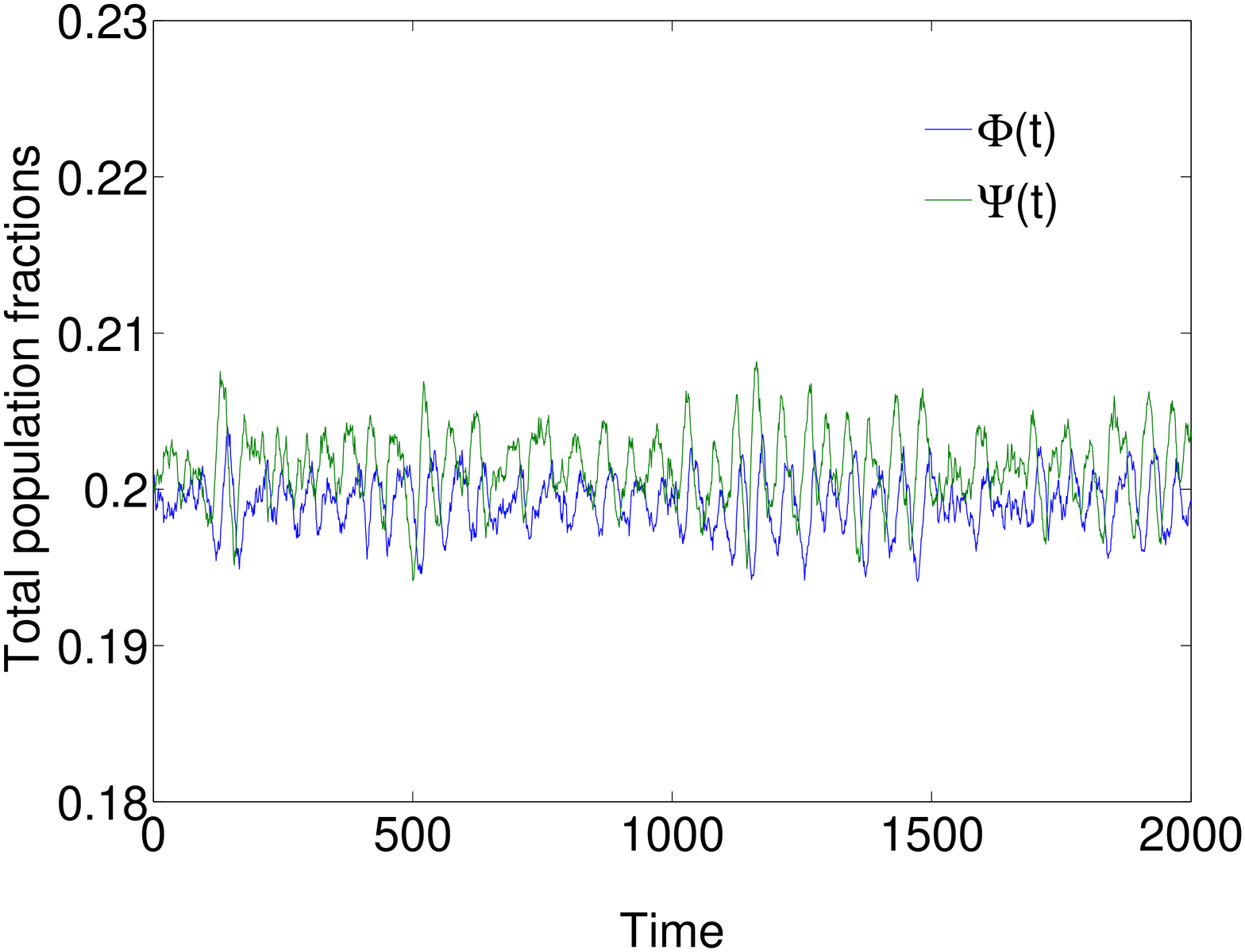}}
\subfigure[]
{\includegraphics[clip,width=1.0\columnwidth,keepaspectratio]{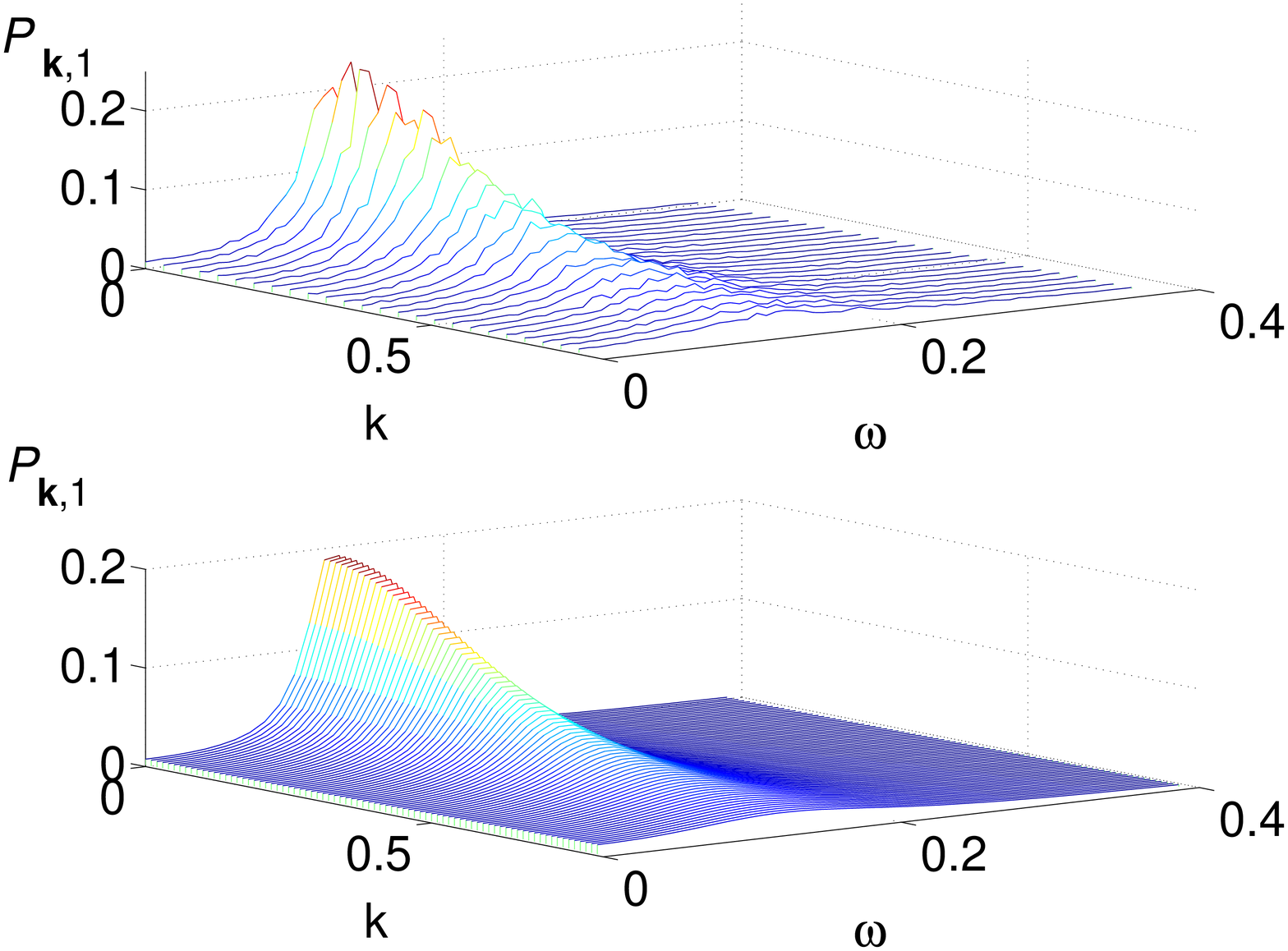}}
\subfigure[]
{\includegraphics[clip,width=1.0\columnwidth,keepaspectratio]{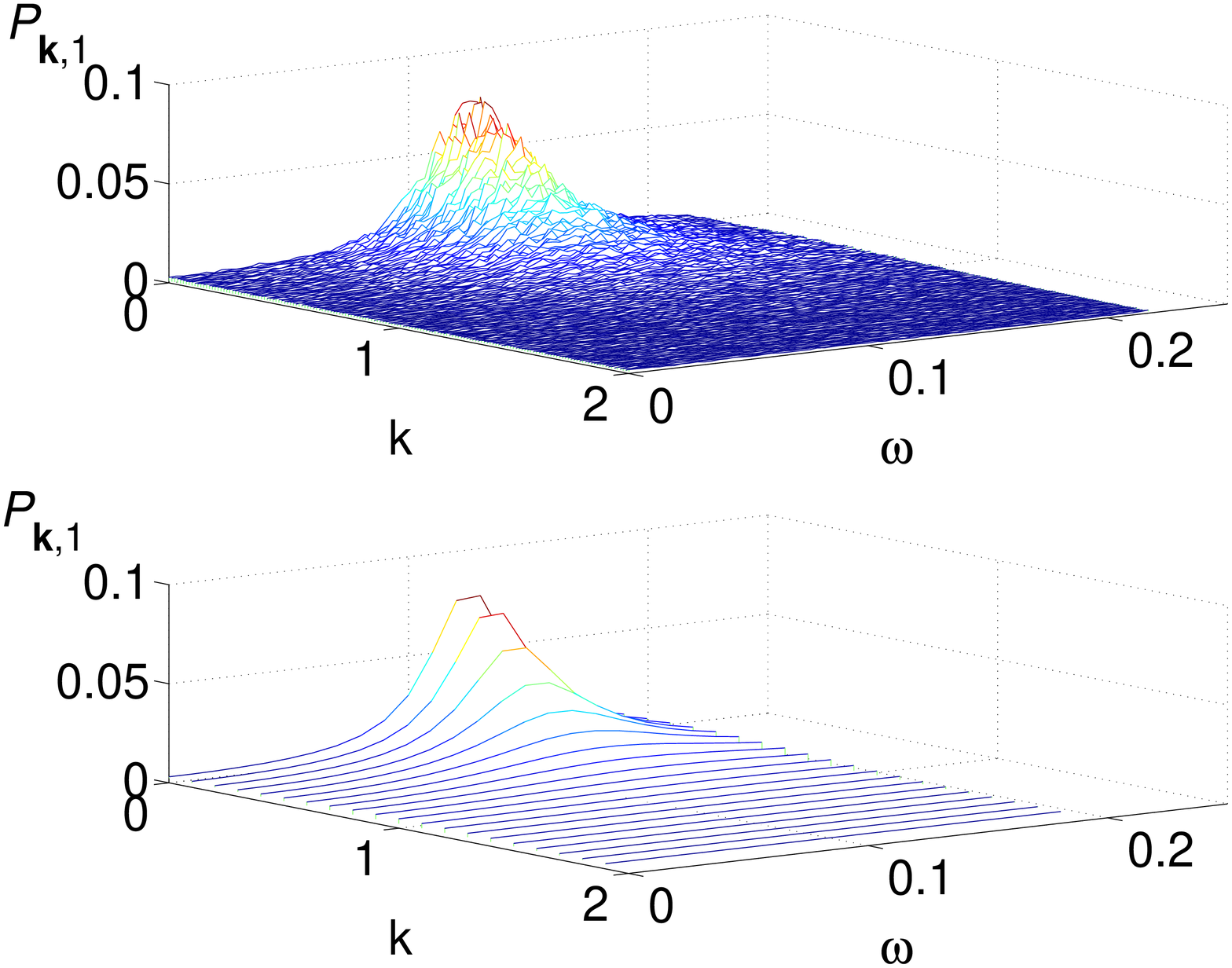}}
\subfigure[]
{\includegraphics[clip,width=1.0\columnwidth,keepaspectratio]{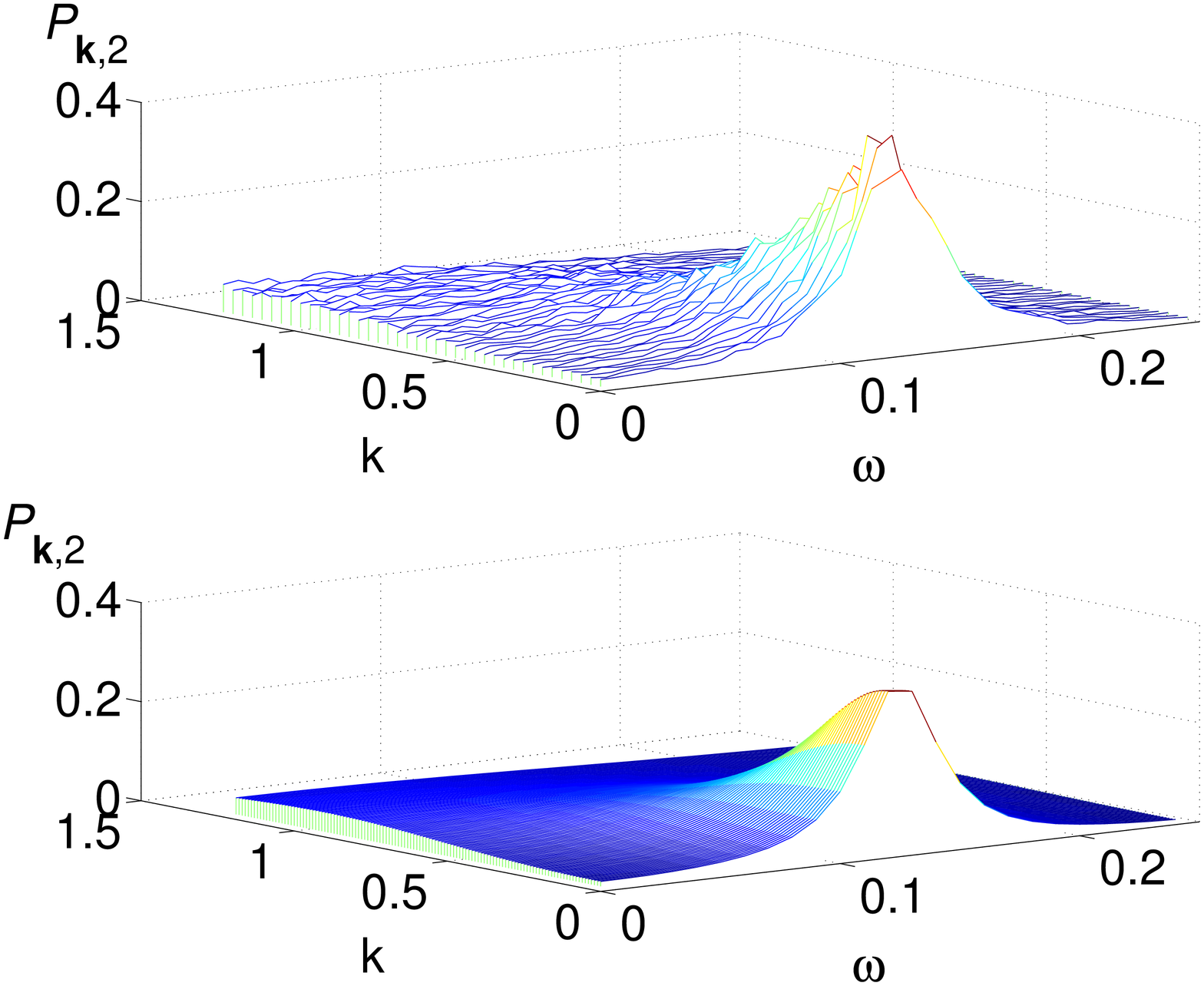}}
\subfigure[]
{\includegraphics[clip,width=1.0\columnwidth,keepaspectratio]{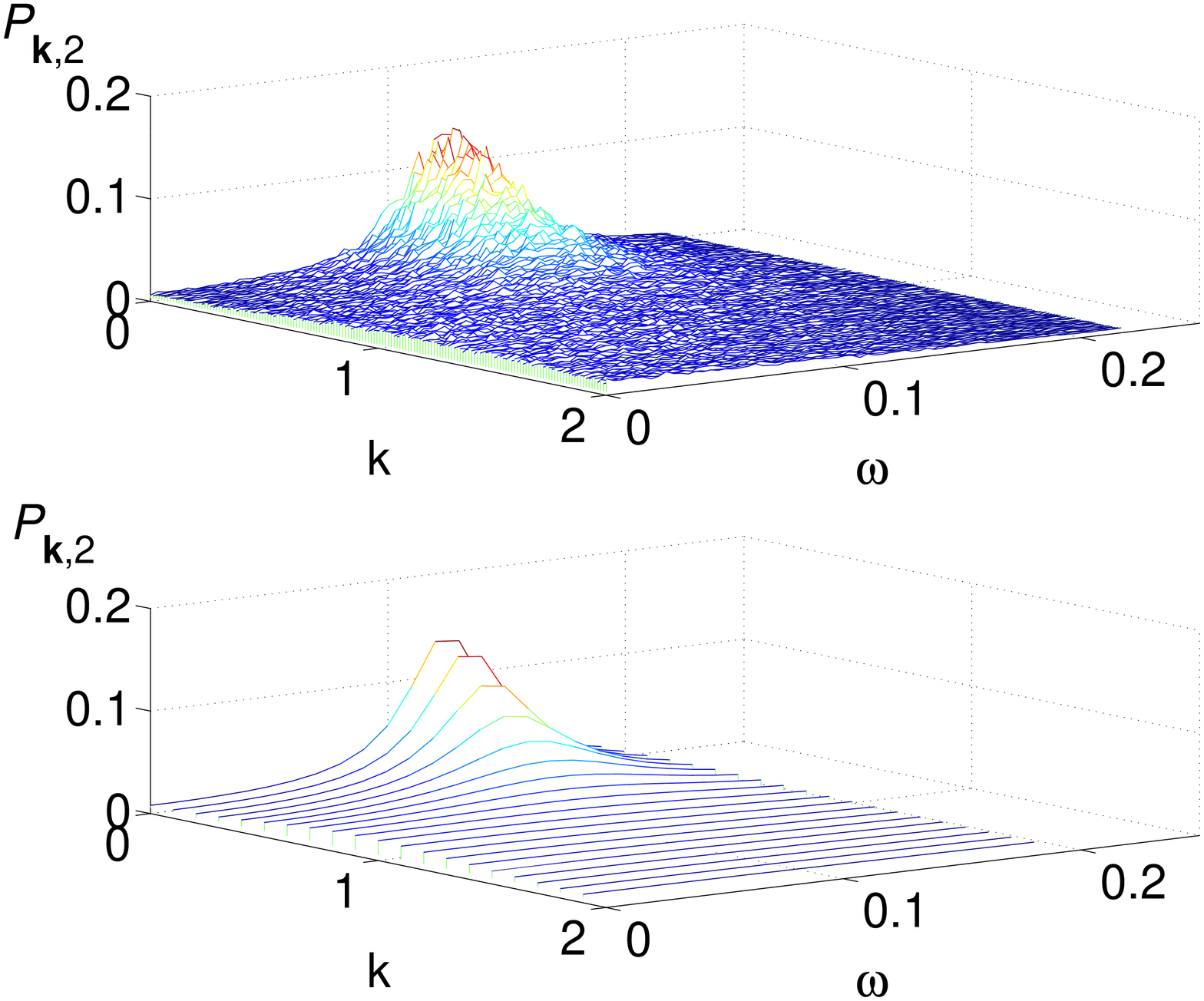}}
\caption{Temporal evolution of the total population fractions and
power spectra obtained from averaging 150 independent realizations
with  $\Omega=200$ (left column), and averaging 100 realizations with
a system composed by $\Omega=500$ sites (right column). The reaction
rates are the same as those indicated in Figure \ref{trun}. The upper
graphs in panels (c)-(f) show the results of the simulations while the 
lower graphs the analytic predictions (\ref{Pwk1})-(\ref{Pwk2}).}
\label{pwks}
\end{figure*}


\begin{figure*}
\centering
\subfigure[]
{\includegraphics[clip,width=1.0\columnwidth,keepaspectratio]{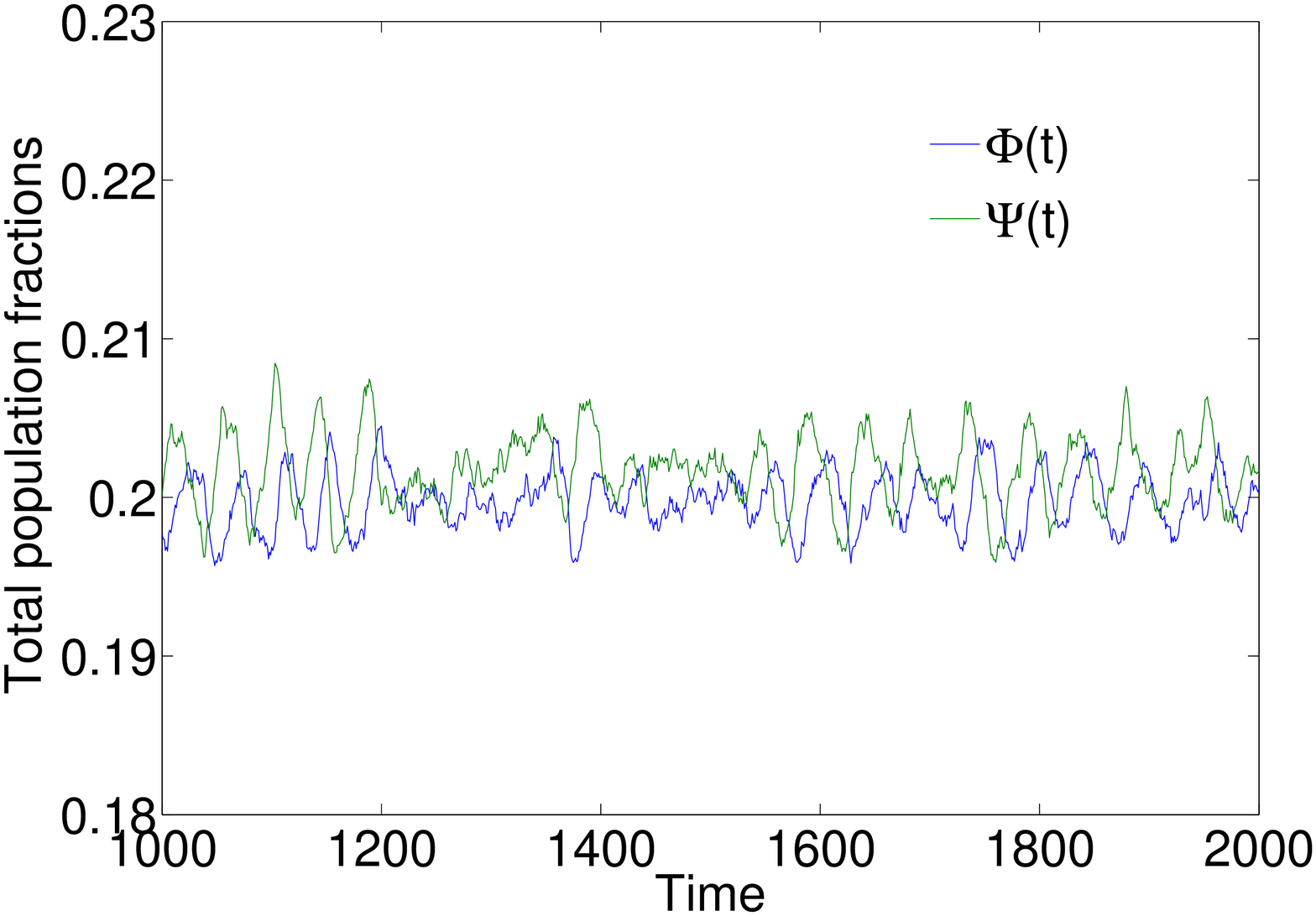}}
\subfigure[]
{\includegraphics[clip,width=1.0\columnwidth,keepaspectratio]{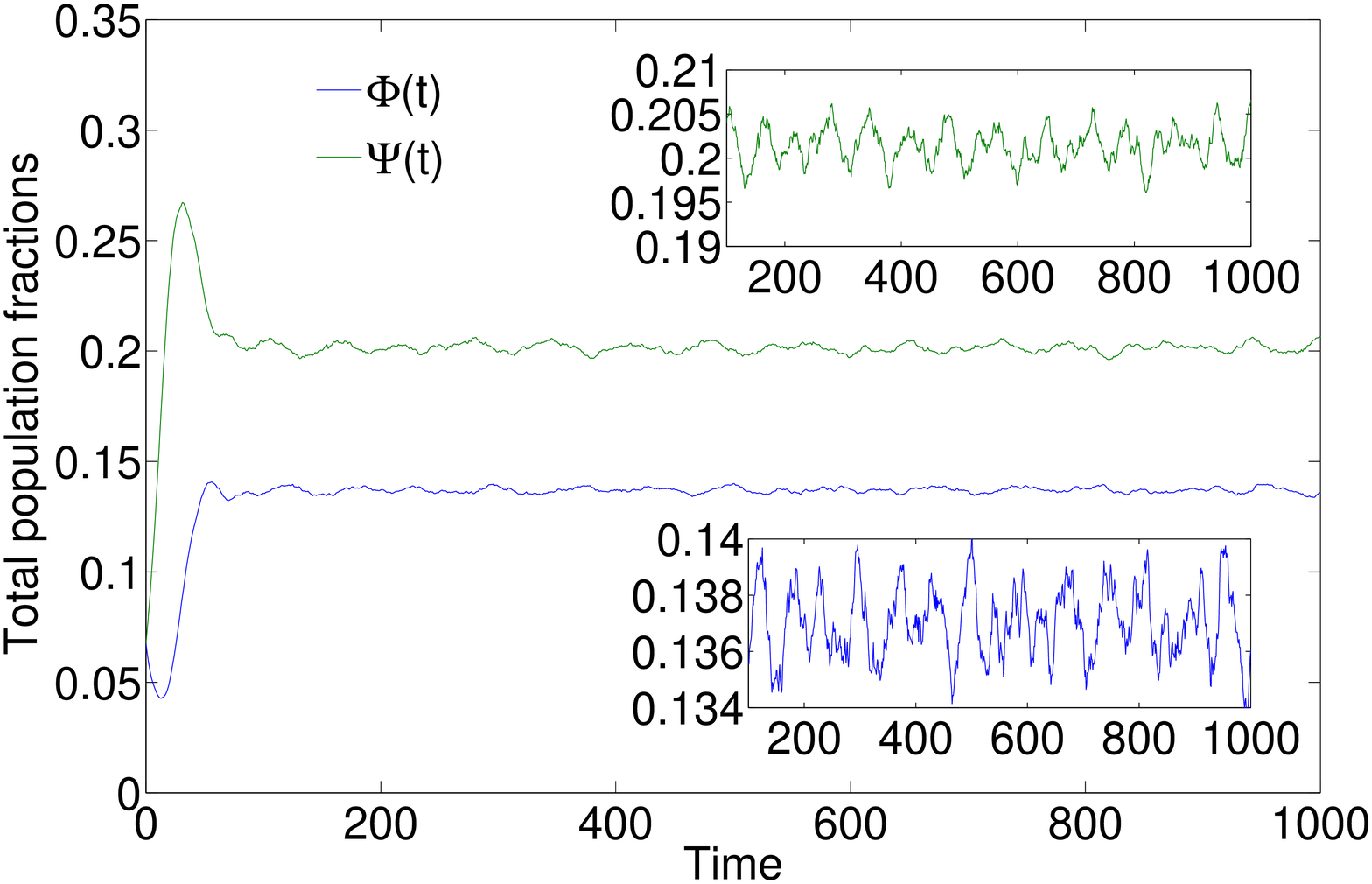}}
\subfigure[]
{\includegraphics[clip,width=1.0\columnwidth,keepaspectratio]{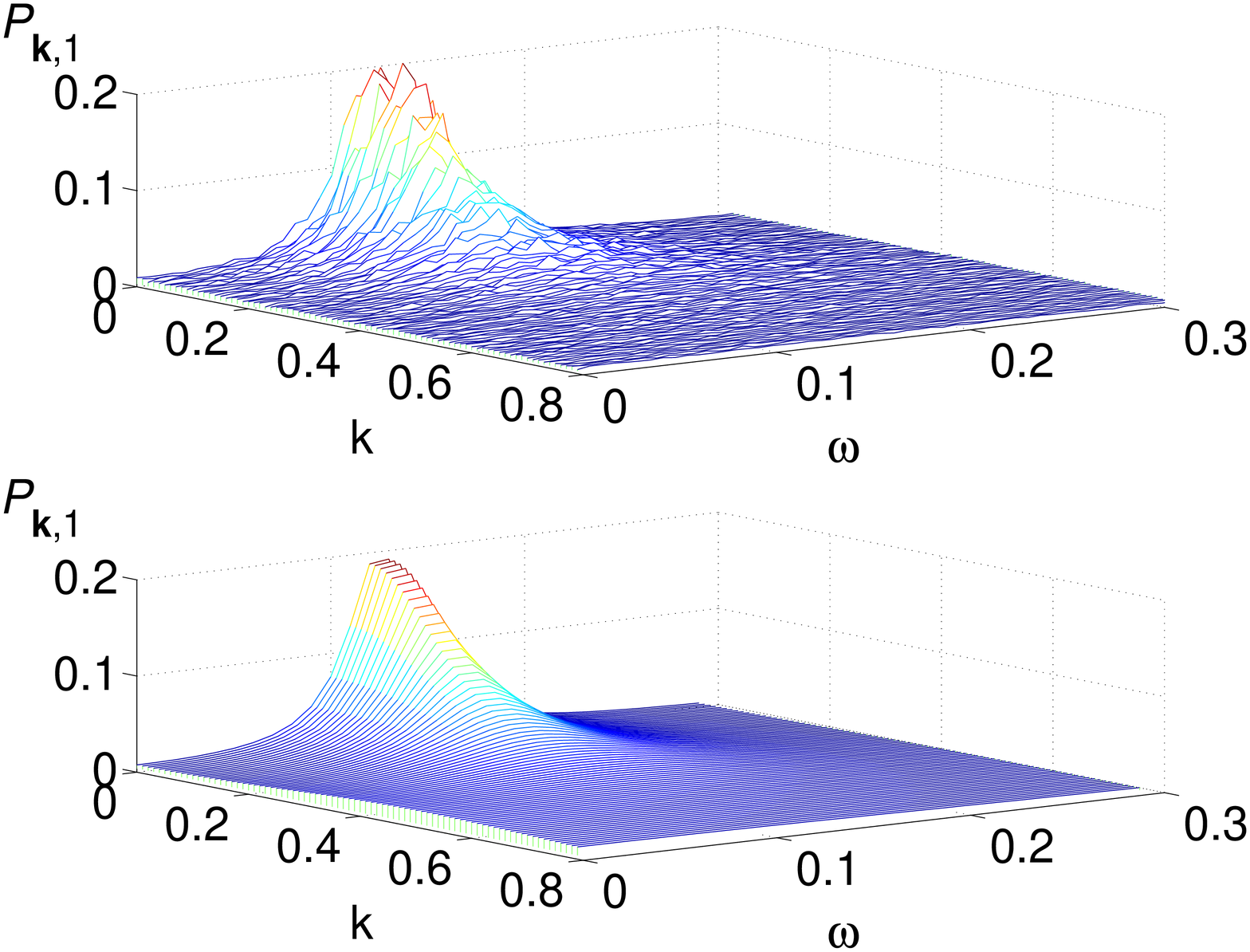}}
\subfigure[]
{\includegraphics[clip,width=1.0\columnwidth,keepaspectratio]{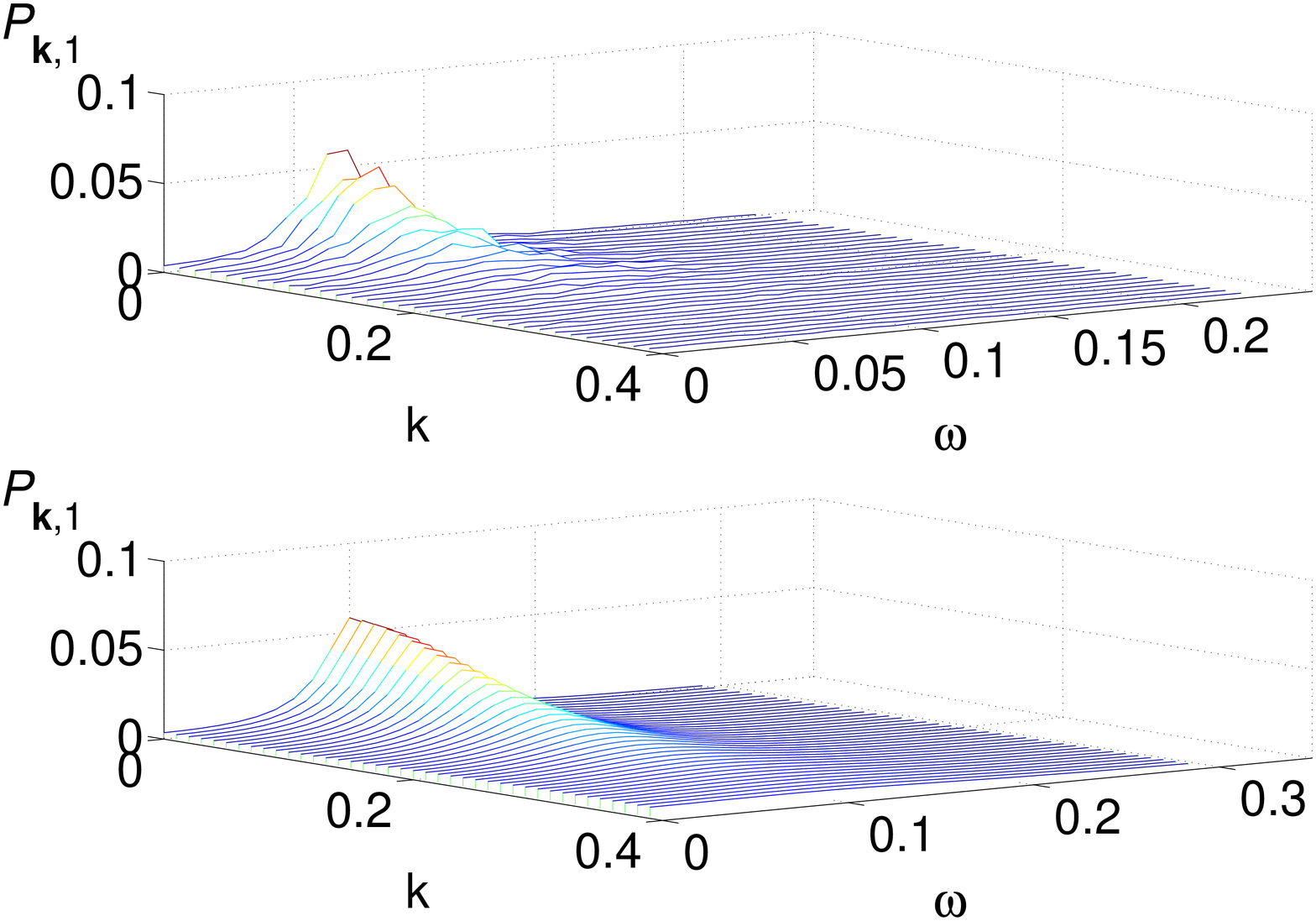}}
\subfigure[]
{\includegraphics[clip,width=1.0\columnwidth,keepaspectratio]{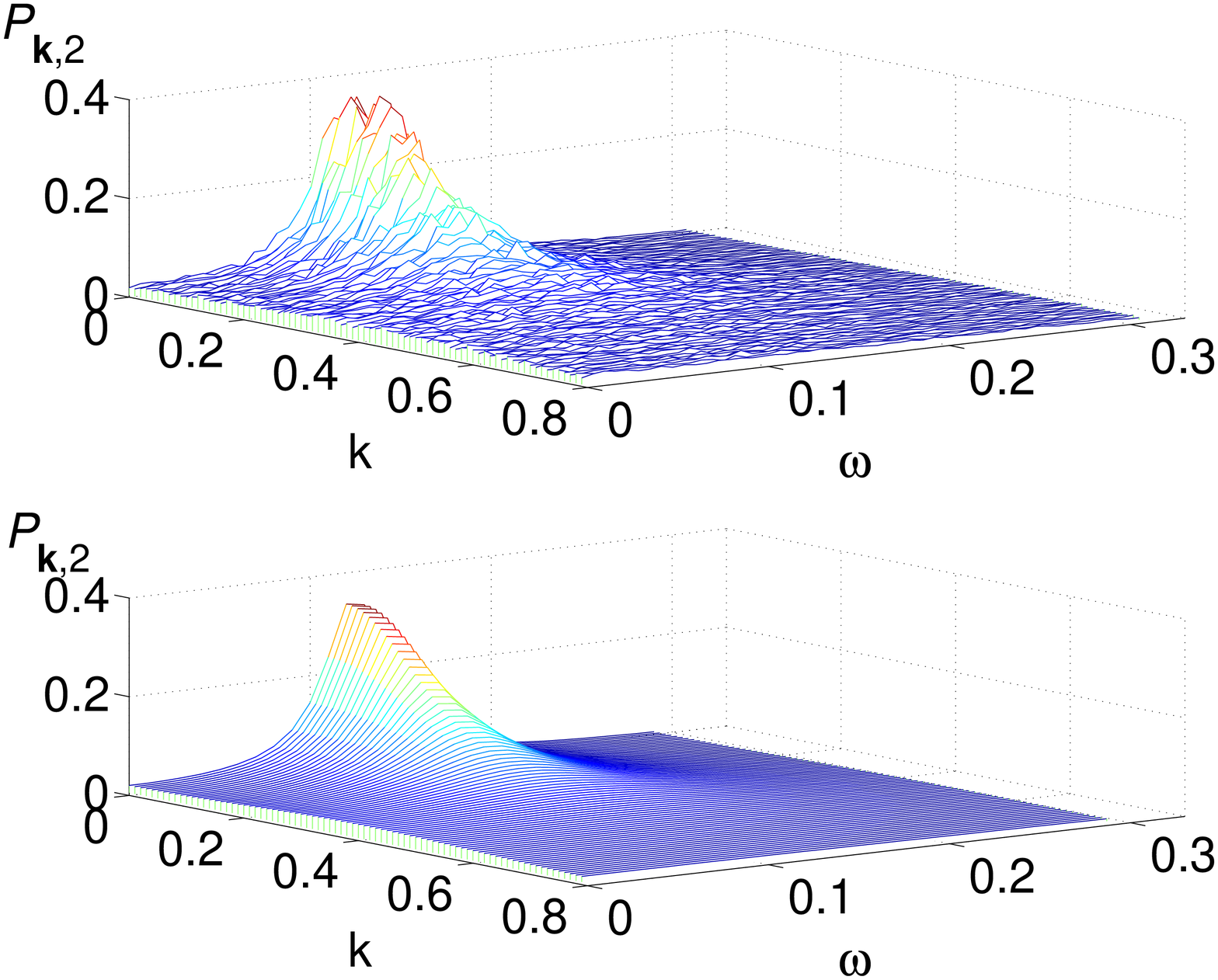}}
\subfigure[]
{\includegraphics[clip,width=1.0\columnwidth,keepaspectratio]{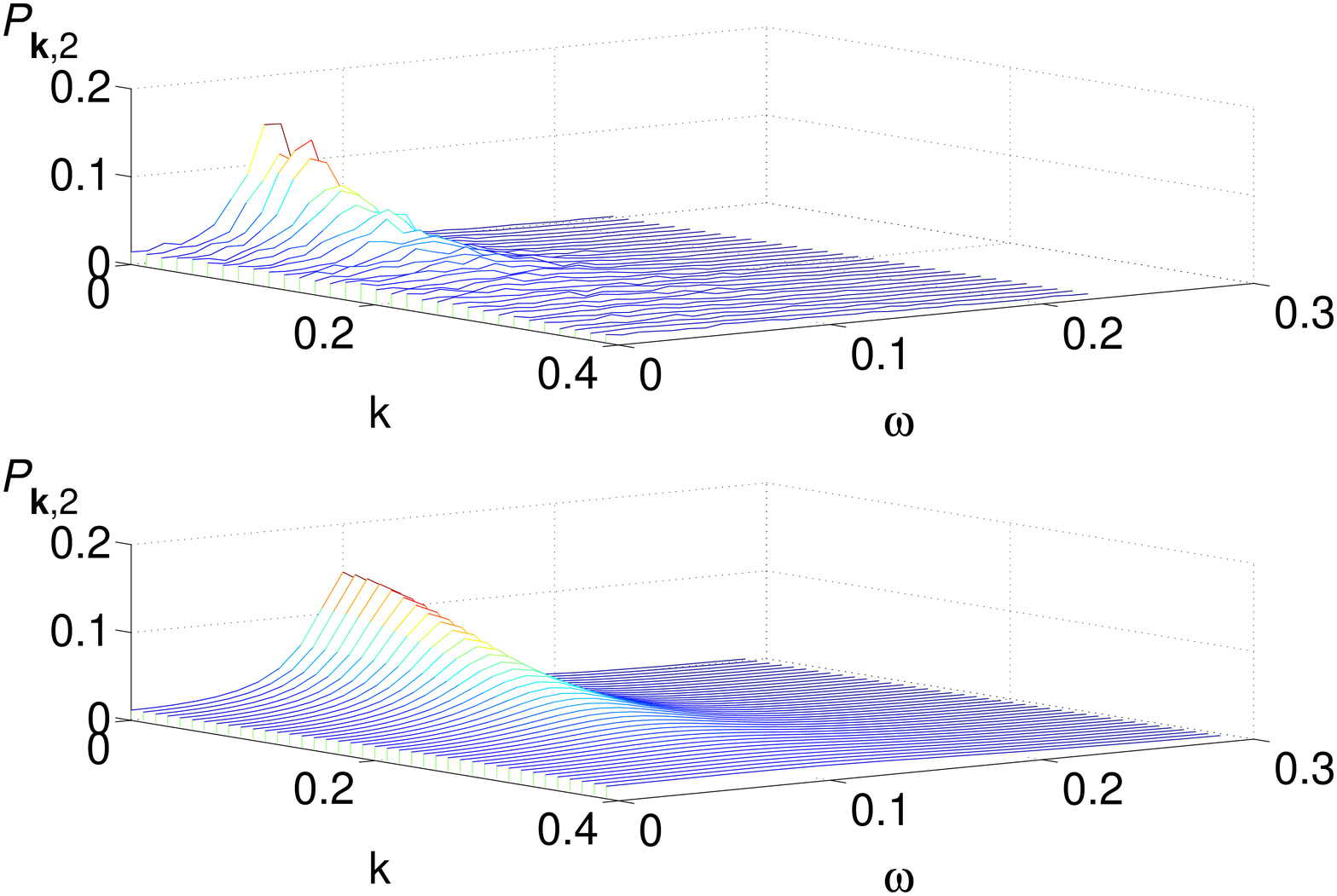}}
\caption{Temporal evolution of the total population fractions and spectra
obtained from numerical simulations of the process (upper graphs) and from
Eqs. (\ref{Pwk1})-(\ref{Pwk2}) (lower graphs). The site capacity and the 
number of sites were $N=500$ and $\Omega=500$. The left-column panels were
obtained employing the same local reaction rates as in the
previous cases and $\mu_1=0.5\Omega$, $\mu_2=0.7\Omega$, whereas the 
right-column panels were obtained with $\mu_1=0.8\Omega$, $\mu_2=0.9\Omega$ 
and $d_2=0.05\Omega$. The spectra in both cases were obtained by averaging 100
independent realizations.}
\label{mporepwks}
\end{figure*}

We used the stochastic simulations of the model defined by 
Eqs.~(\ref{spr1})-(\ref{spm}) using the Gillespie algorithm\,\cite{DTG76}, 
already mentioned in Section \ref{sims}, to determine the Fourier 
transform of the fluctuations $\xi_{k}(\omega)$ and $\eta_{k}(\omega)$. These
were then compared to those from the power spectra (\ref{pred_PS}) and 
(\ref{prey_PS}). Once again we restricted ourselves to one dimension, which 
enabled us to obtain quite comprehensive results. In practice the Fourier
transforms are calculated by employing a discrete Fourier transform, and in 
order to compare the amplitudes obtained numerically with the analytical 
results, the numerically averaged spectra contain an extra factor 
$\left|(4\delta_x\delta_t)/({\cal N}_x{\cal N}_t)\right|^2$, where the 
$\delta$ are the spacing between consecutive points and the ${\cal N}$ the 
number of sampled points in space and time.

We begin by showing the results of changing the number of sites, $\Omega$. In
Fig.~\ref{pwks} the left-hand column shows results obtained by taking 
$\Omega=200$ with all other parameters taking on the same values as in 
Section \ref{sims}. The right-hand column shows results with the same 
parameters again, except that $\Omega=500$. The results from simulations 
were obtained by averaging $100$ realizations of the process, taking an 
initial configuration to be the stationary state in the entire interval,
and only once the oscillatory regime had been established. Specifically 
simulation times were in the interval $t\in\left[1000,2000\right]$. 

The first two figures (\ref{pwks}(a) and (b)) show the typical temporal 
evolution of the total population fractions. Subsequently, the results of 
simulations (upper graphs of Figs.~\ref{pwks}(c)-(f)) and the analytic 
expressions (\ref{Pwk1}) and (\ref{Pwk2}) (lower graphs of 
Figs.~\ref{pwks}(c)-(f)) are displayed. Mention should be made of the 
scales of these (and subsequent) figures. The $k$ take on discrete 
values $2\pi n$ where $n$ is an integer, since the length of the interval 
being considered is unity. In order to compare to the analytic forms, $k$ 
is measured in units of $1/a$, and so effectively it is $ak$ which is 
plotted. This takes on discrete values $2\pi n/\Omega$, but we are looking 
at sufficiently large values of $\Omega$ that the $k$ values appear 
continuous. For the $\omega-$axis, the characteristic time which sets the 
scale is $\delta_{t}$. It should also be noted that the $k$ axis in 
Fig.~\ref{pwks}(e) has been reversed to show the peak from another 
perspective. From Fig.~\ref{pwks}(d) and (f), we see that the predator and 
prey spectra do not seem to differ appreciably. This was also found in the 
non-spatial case\,\cite{AJMaTJN05}. However, as we shall see later, if 
the migration rates are significantly different then the two spectra will 
differ. Also the fact that $\alpha_{{\bf k}, 11} \neq 0$, but that the 
analogous quantity in the non-spatial case, $a_{11}$, does vanish, leads 
to additional differences between the predator and prey spectra in the 
spatial version.

For both values of $\Omega$ studied, we observe that the analytic expressions 
and those obtained from simulating the full stochastic process show good 
agreement, which indicates that the use of the first two orders in the van 
Kampen approximation are sufficient for our purposes. We see that there is a 
large peak at a non-zero value of $\omega$ and so resonant behavior still 
occurs in this spatial model, just as it did in the non-spatial case. However,
the height of the peak reduces with $k$ and eventually at some finite value of
$k$ the peak disappears altogether. There is always an additional peak at
$\omega=0$; this is much smaller and is just visible in Figs.~\ref{pwks}(e)
and (f). We will discuss it again shortly, when a different choice of the 
migration rates makes it far more prominent.  

In Fig.~\ref{mporepwks} similar plots are shown for two different values of
the migration rates $\mu_1$ and $\mu_2$, keeping all other parameters as 
before (except in one case where we take $d_{2} \neq 0$) and taking 
$\Omega=500$. The value of $d_2$ was changed so that the fixed-point values
$\phi^*$ and $\psi^*$ were different, which made some of the plots clearer.

\begin{figure*}
\centering 
\subfigure[]
{\includegraphics[clip,width=1.0\columnwidth,keepaspectratio]{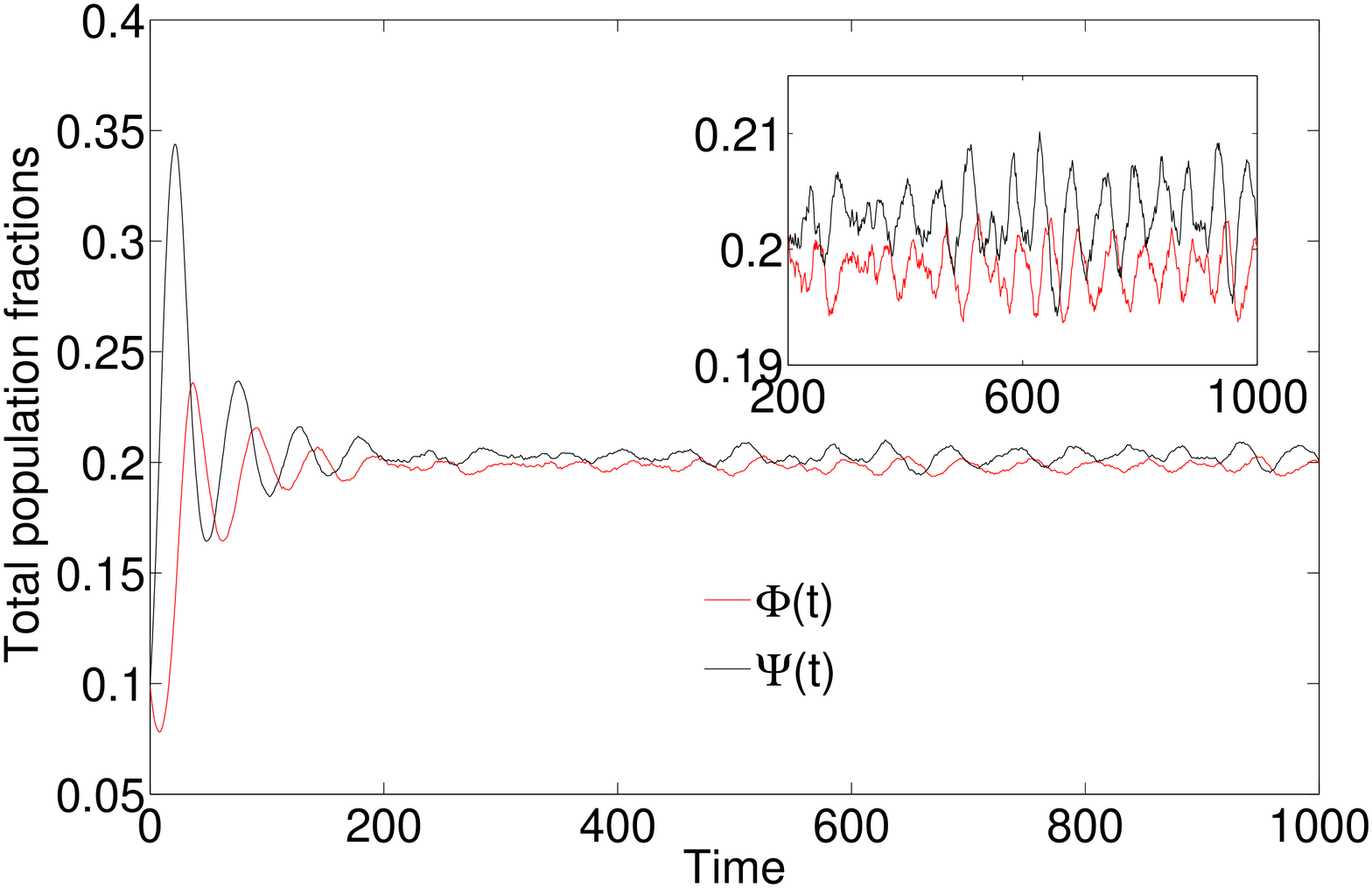}}
\subfigure[]
{\includegraphics[clip,width=1.0\columnwidth,keepaspectratio]{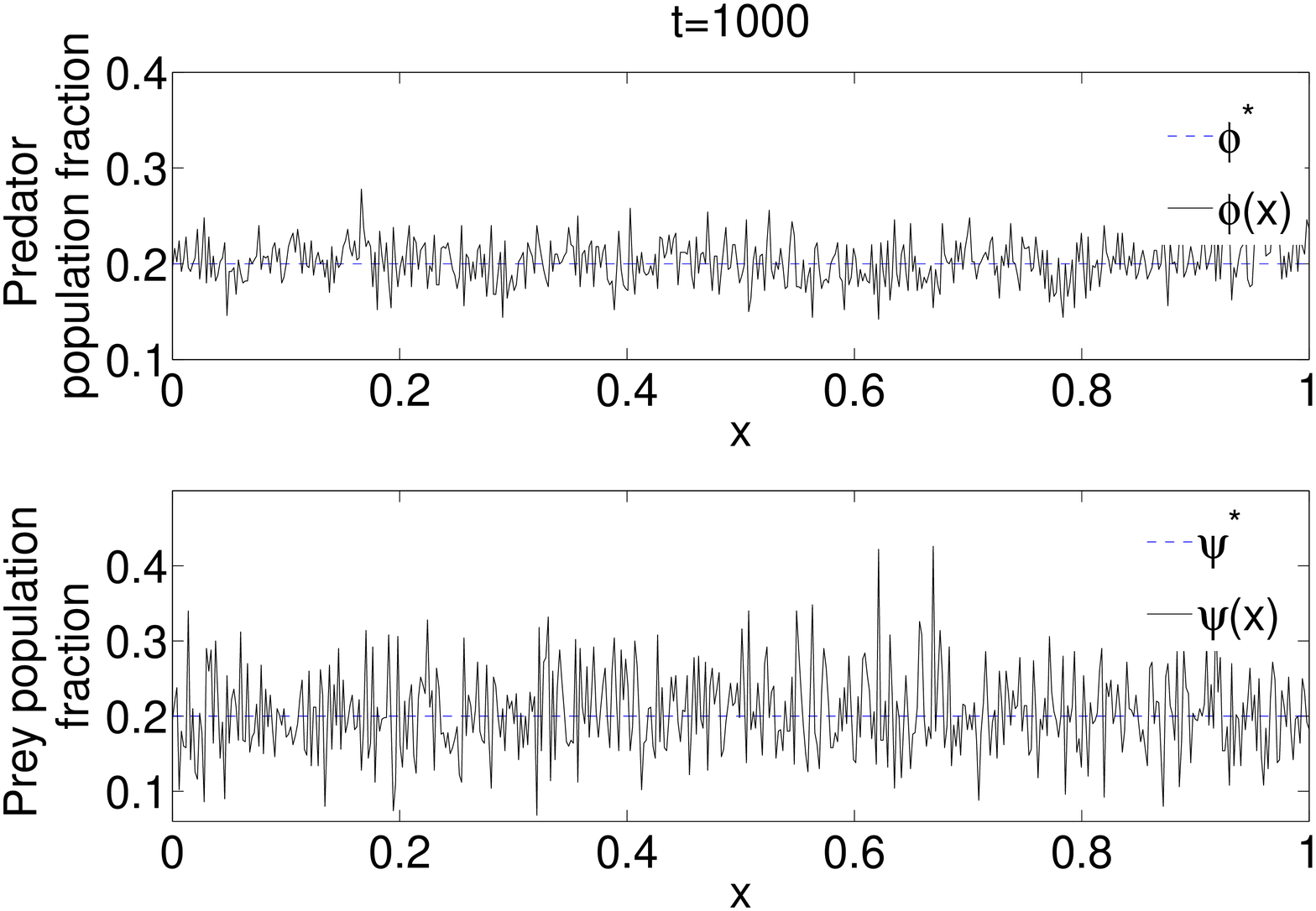}}
\subfigure[]
{\includegraphics[clip,width=1.0\columnwidth,keepaspectratio]{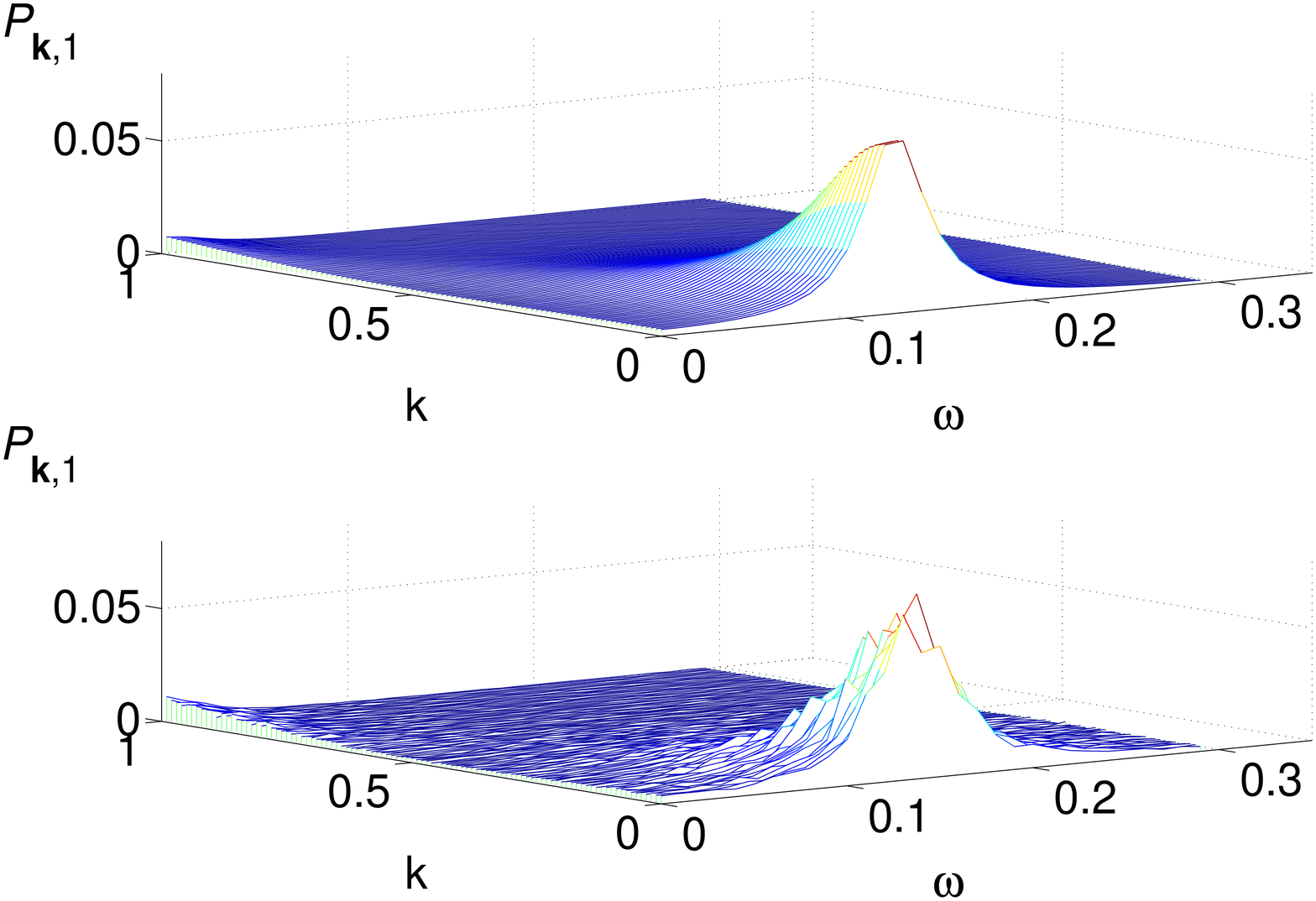}} 
\subfigure[]
{\includegraphics[clip,width=1.0\columnwidth,keepaspectratio]{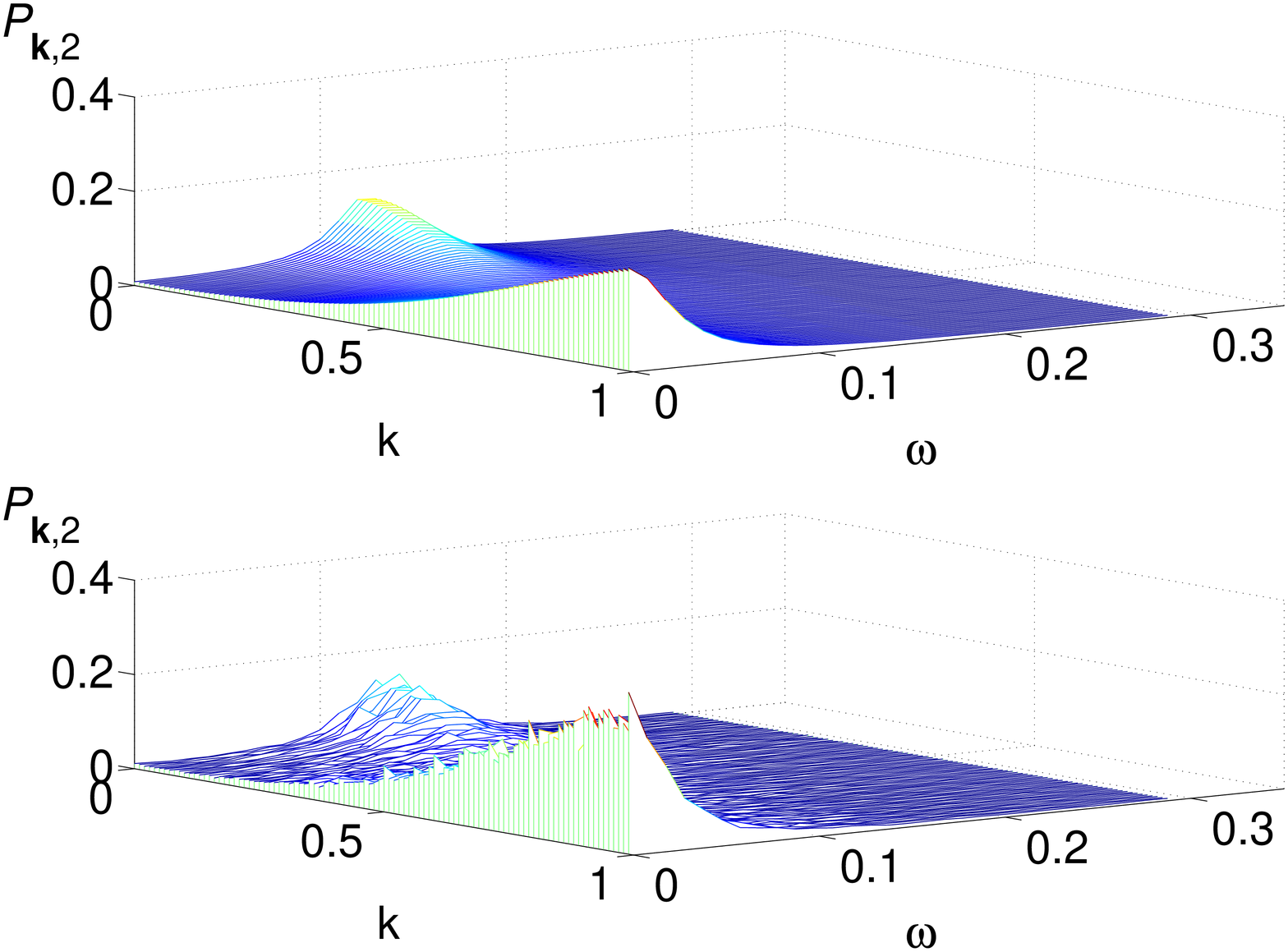}}
\caption{(a) Total population fractions and (b) spatial configurations
for the predator and prey fractions. (c)-(d) Numerically and
analytically obtained power spectra obtained from $70$
realizations of the process and from expressions (\ref{Pwk1}) and
(\ref{Pwk2}) respectively. The migration rates were
$\mu_1=1.0\Omega$, $\mu_2=0.01\Omega$, and the local rates are the same
as in Figure \ref{trun}. The amplification effect is stronger that
in the previous cases particularly in the  case of the prey
spectra. Simulations have been carried out swapping the values of the
rates, showing a similar effect, but for the other spectrum.}
\label{bigm}
\end{figure*}

Finally, as shown in Fig.~\ref{bigm}, we found that making one migration 
rate considerably bigger than the other led to significant differences. 
Although the peaks at non-zero $\omega$ were still present, they looked 
rather different for the predator and for the prey spectra. Also noteworthy 
is the peak at zero frequency, which is now much larger than before in the 
case of the prey. The graph is cut-off at $k \sim 1$ only because it 
becomes much more noisy at larger values of $k$ and so rather difficult to 
interpret. A similar result is obtained if we swap the values of the 
migration rates, but now it will be the predator fluctuations which will 
exhibit the large amplification effect.
 

\section{Conclusion}
\label{conclude}
In the work that we have presented here we have stressed the systematic nature 
of the procedures employed and the generic nature of the results obtained. The
starting point was the ILM (\ref{spr1})-(\ref{spm}), but many of the results 
that we give are not sensitive to the precise form of the model employed. For 
instance, births and predator events could have an alternative (or additional) 
rule which would involve nearest-neighbor patches. An example would be 
$B_{i} E_{j} \rightarrow B_{i} B_{j}$, where $i$ and $j$ are nearest 
neighbor sites, which would mean that a birth could only take place of there 
was space in the adjoining patch. The definition of the neighborhood could 
also vary to include next-nearest neighbors or a Moore neighborhood, rather 
than a von Neumann one. All these changes would give the same behavior at the 
population level, and in many cases exactly the same model, and leave the form 
of our results unchanged. 

In a similar way, the nature of the lattice, and its dimension, only enter 
the differential equations through the discrete Laplacian operator 
$\Delta_{\bf k}$ and factors of $a^d$, leaving the essential aspects of 
quantities such as the power spectra unchanged. One consequence of this 
observation is that the very good agreement between the analytically 
calculated power spectra and those found from the one-dimensional 
simulations should still occur in higher dimensions and for other models. 
This is the main justification for restricting our simulations to one 
dimension and hence being able to obtain higher quality data. All these 
observations lead us to expect our results to be generally applicable and to 
be capable of straightforward generalization to other, similar, problems.

The procedure we have followed is also systematic. Rather than writing down 
a PLM on phenomenological grounds, we have derived it within a expansion 
procedure with a small parameter ($1/\sqrt{N}$) from a more basic ILM. This 
allows us to relate the parameters of the PLM to those of the ILM, but also 
to derive the strength and nature of the noise that is a manifestation of the 
demographic stochasticity, rather than putting it in by hand. The two sets 
of equations derived from the ILM --- the macroscopic, or mean-field equations 
and the Langevin equations describing the stochastic fluctuations about the 
mean fields --- capture the essential aspects of the dynamics at the 
population level. Provided that $N$ is not too small that stochastic extinction
events are significant, they give a very good description of generic phenomena 
which one would expect to see in simple descriptions of systems with one 
predator species and one prey species.   

The main focus of this paper was on the power spectra. We found that the
resonant amplification present in the well-mixed system is still present in
the spatial system, although the height of the peak decreases with $k$, at
least in the one-dimensional model. The spectra for the predator and prey 
species can be made significantly different by making one of the migration 
rates much bigger than the other, a freedom that was not available to us in 
the non-spatial case. There is also a peak at $\omega=0$. This is present 
in the non-spatial model, but has no physical significance. Here it does: it 
corresponds to periodic spatial structures. This peak is very small if the
migration rates are of the same order, but can be as large as the peak at 
$\omega \neq 0$ if the migration rates are sufficiently different.

The existence of a large peak at non-zero $\omega$ and $|{\bf k}|$ means that
when the system is studied at a spatial resolution defined by ${\bf k}$, there
will large amplitude oscillations of frequency $\omega_{0}({\bf k})$, where
this is the position of the peak. While we can deduce the existence of 
such structures for general $d$ from our analytic calculations, our numerical
work has only been undertaken for $d=1$. Since the topology of one-dimensional 
lattices constrain the dynamics from exhibiting more interesting structures 
in space and time (as have been reported in numerical studies of models of a 
similar nature\,\cite{WGW93,MMITG07,JES96}), these periodic structures may 
have more complicated forms in higher dimensions. 

The approach which consists of defining the time-evolution of a model by a 
master equation, and then performing some type of analysis which allows one to 
obtain not only the mean field theory, but corrections to it, has proved to be 
very effective in understanding the results obtained from numerical 
simulations\,\cite{OOSJC06,JES96,mck07}. In the case of the technique employed 
in this paper, there are many applications which can be envisaged --- those
which apply to completely different systems, but also predator-prey systems 
with a more complicated functional response. It would also be interesting 
to investigate systems whose deterministic limit exhibits Turing 
instabilities\,\cite{MP93,DAFB02}. In other words, the general approach 
we have discussed here, and the results we have reported, have a very general
nature. This implies that resonant amplification of stochastic fluctuations 
will be frequently seen in lattice models and lead to cyclic behavior in a 
wide range of systems. 

\begin{acknowledgments}
We thank Andrew Black and Tobias Galla for useful discussions. CAL acknowledges
the award of a studentship from CONACYT (Mexico) and AJM of a grant 
(GR/T11784/0) from the EPSRC (UK). 

\end{acknowledgments}

\appendix

\section{System size expansion}
\label{VK}

In this Appendix the master equation for the model discussed in the main text
is expanded to leading order (which gives the macroscopic laws) and 
next-to-leading order (which gives the linear noise approximations) in the
van Kampen system-size expansion\,\cite{NGvK92}. The system-size expansion is
not usually applied to systems with spatial degrees of freedom (but see 
\cite{her90}), and there are a number of possible ways of proceeding. Here we 
will take what is perhaps the simplest case, and assume that the expansion 
parameter is $1/\sqrt{N}$, that is, each lattice site is treated as a subsystem
for which the carrying capacity becomes large. The calculation may be 
performed in a way which is similar to the non-spatial case; whereas in the 
non-spatial model there were two degrees of freedom: the number of predators, 
$n$, and the number of prey, $m$, there are now $2\Omega$ degrees of freedom, 
$n_i$ and $m_i$, $i=1,\ldots,\Omega$. In what follows we will therefore limit 
ourselves to an outline of the method and to the statement of key intermediate 
results. For all fuller description of the method, reference should be made 
to van Kampen's book\,\cite{NGvK92} or papers which apply the method to 
related problems\,\cite{alo07,mck07}.

The system-size expansion begins with the mapping
\begin{equation} 
\frac{n_{i}}{N}=\phi_{i}+(N)^{-\frac{1}{2}}\xi_{i}\,, \ \ 
\frac{m_{i}}{N}=\psi_{i}+(N)^{-\frac{1}{2}}\eta_{i}\,.
\label{vK_ansatz}
\end{equation}
Here $\phi_{i}(t)$ and $\psi_{i}(t)$ will be the variables in the PLM, and
the stochastic variables $\xi_{i}(t)$ and $\eta_{i}(t)$ will appear in the 
Langevin equations at next to leading order.

Under this transformation, the left-hand side of the master 
equation (\ref{master}) becomes:
\begin{equation}
\frac{\partial\Pi}{\partial t}+\sum_{i=1}^{\Omega}\left(\dot{\xi}_{i}
\frac{\partial\Pi}{\partial\xi_{i}}+\dot{\eta}_{i}\frac{\partial\Pi}
{\partial\eta_{i}}\right)\,,
\label{LHS}
\end{equation}
where $\dot{\xi}_{i}=-(N)^{\frac{1}{2}}\dot{\phi}_{i}$, 
$\dot{\eta}_{i}=-(N)^{\frac{1}{2}}\dot{\psi}_{i}$ and where $\Pi$ is 
the probability density function, but now expressed as a function of 
$\phi_{i}, \psi_{i}$ and $t$. To determine the form of the right-hand side of 
the master equation in terms of the new variables, we need to write 
${\cal T}^{\rm loc}_{i}$ and ${\cal T}^{\rm mig}_{ij}$, given by 
Eqs.~(\ref{local_contrib}) and (\ref{mig_contrib}) respectively, in terms of 
these new variables. This consists of two stages: first writing the step 
operators (\ref{step}) as operators involving the new variables, and secondly, 
determining their action on the transition probabilities (\ref{pr}) and 
(\ref{nn}).

Beginning with ${\cal T}^{\rm loc}_{i}$ the first stage gives
\begin{eqnarray}
E_{x_i}-1 & = & N^{-\frac{1}{2}}\frac{\partial}{\partial\xi_i}+
\frac{1}{2}N^{-1}\frac{\partial^{2}}{\partial\xi_i^{2}}+\ldots\nonumber\\
E_{y_i}-1 & = & N^{-\frac{1}{2}}\frac{\partial}{\partial\eta_i}+
\frac{1}{2}N^{-1}\frac{\partial^{2}}{\partial\eta_i^{2}}+\ldots\nonumber\\
E_{x_i}^{-1}-1 & = & -N^{-\frac{1}{2}}\frac{\partial}{\partial\xi_i}+
\frac{1}{2}N^{-1}\frac{\partial^{2}}{\partial\xi_i^{2}}+\ldots\nonumber\\
E_{x_i}^{-1}E_{y_i}-1 & = & N^{-\frac{1}{2}}\frac{\partial}{\partial\eta_i}
-N^{-\frac{1}{2}}\frac{\partial}{\partial\xi_i}\nonumber\\
& &+\frac{1}{2}N^{-1}\left(\frac{\partial}{\partial\xi_i}
-\frac{\partial}{\partial\eta_i}\right)^{2}+\ldots\,.\label{E-1}
\end{eqnarray}
We can now list the various contributions we obtain, at order $N^{\frac{1}{2}}$
and $N^{0}$, which we need in order to find ${\cal T}^{\rm loc}_{i}$ as defined
in Eq.~(\ref{local_contrib}):
\begin{itemize}
\item[(i)] $\left(E_{x_i}-1\right)d_{1}n_i$:
\begin{eqnarray}
N^{\frac{1}{2}}&:&d_{1}\phi_i\frac{\partial}{\partial\xi_i}\nonumber\\
N^{0}&:&d_{1}\frac{\partial}{\partial\xi_i}\xi_i,\:
\frac{1}{2}d_{1}\phi_i\frac{\partial^{2}}{\partial\xi_i^{2}}\nonumber
\end{eqnarray}

\item[(ii)] $\left(E_{y_i}-1\right)\left(\frac{2p_{2}n_im_i}{N}+
d_{2}m_i\right)$:
\begin{eqnarray}
N^{\frac{1}{2}}&:& \left(d_{2}\psi_i+2p_{2}\psi_i\phi_i\right)
\frac{\partial}{\partial\eta_i}\nonumber\\
N^{0}&:&\left(d_{2}+2p_{2}\phi_i\right)\frac{\partial}{\partial\eta_i}
\eta_i\nonumber,\:
2p_{2}\psi_i\frac{\partial}{\partial\eta_i}\xi_i,\nonumber\\ 
& &\left(\frac{d_{2}}{2}\psi_i+p_{2}\psi_i\phi_i\right)
\frac{\partial^{2}}{\partial\eta_i^{2}}\nonumber
\end{eqnarray}

\item[(iii)] $\left(E_{y_i}^{-1}-1\right)
\left(\frac{2bm_i(N-n_i-m_i)}{N}\right)$:
\begin{eqnarray}
N^{\frac{1}{2}}&:& -2b\psi_i\left(1-\psi_i-\phi_i\right)
\frac{\partial}{\partial\eta_i}\nonumber\\
N^{0}&:&2b\left(2\psi_i-1+\phi_i\right)\frac{\partial}{\partial\eta_i}
\eta_i\nonumber\\
& &2b\psi_i\frac{\partial}{\partial\eta_i}\xi_i,\: b\psi_i
\left(1-\psi_i-\phi_i\right)\frac{\partial^{2}}{\partial\eta_i^{2}}\nonumber
\end{eqnarray}

\item[(iv)] $\left(E_{x_i}^{-1}E_{y_i}-1\right)
\left(\frac{2p_{1}n_im_i}{N}\right)$:
\begin{eqnarray}
N^{\frac{1}{2}}&:&-2p_{1}\psi_i\phi_i\frac{\partial}{\partial\xi_i},\:
2p_{1}\phi_i\psi_i\frac{\partial}{\partial\eta_i}\nonumber\\
N^{0}&:& 2p_{1}\phi_i\frac{\partial}{\partial\eta_i}
\eta_i,\: 2p_{1}\psi_i\frac{\partial}{\partial\eta_i}\xi_i,\:
p_{1}\phi_i\psi_i\frac{\partial^{2}}{\partial\eta_i^{2}},\: 
p_{1}\phi_i\psi_i\frac{\partial^{2}}{\partial\xi_i^{2}},\nonumber\\
& &-2p_{1}\phi_i\frac{\partial}{\partial\xi_i}\eta_i,\
: -2p_{1}\psi_i\frac{\partial}{\partial\xi_i}\xi_i,\:
-2p_{1}\phi_i\psi_i\frac{\partial^{2}}{\partial\eta_i\partial\xi_i}\,.
\nonumber
\end{eqnarray}
\end{itemize}
Identifying the terms of order $N^{\frac{1}{2}}$ on the right- and left-hand 
sides of the master equation gives the contributions of the local reactions 
to the macroscopic laws:
\begin{eqnarray}
-\dot{\phi_{i}} &=& \frac{d_{1}}{\Omega}\phi_{i}-
\frac{2p_{1}}{\Omega}\psi_{i}\phi_{i}, \label{local_macro_1} \\
-\dot{\psi_{i}} &=& \frac{d_{2}}{\Omega}\psi_{i}+
\frac{2p_{2}}{\Omega}\psi_{i}\phi_{i}
+\frac{2p_{1}}{\Omega}\phi_{i}\psi_{i} \nonumber \\
&-& \frac{2b}{\Omega}\psi_{i} \left(1-\psi_{i}-\phi_{i}\right)\,.
\label{local_macro_2}
\end{eqnarray}
If a rescaled time, $\tau = t/\Omega$, is introduced, then these equations 
are exactly the PLM of the non-spatial version of the model\,\cite{AJMaTJN05}.
This is as it should be, since without including the nearest-neighbor couplings
in ${\cal T}^{\rm mig}_{ij}$, the system is simply $\Omega$ copies of the 
non-spatial model.

Performing a similar identification of both sides of the master equation, but 
now for terms of order $N^{0}$ gives a Fokker-Planck equation:
\begin{equation}
\frac{\partial \Pi}{\partial t} = - \sum^{\Omega}_{i=1} \frac{\partial}
{\partial \boldsymbol{\zeta}_i} \left[ {\cal A}_{i}(\boldsymbol{\zeta}(t)) 
\Pi \right] + \frac{1}{2} \sum_{i,j} \frac{\partial^2}
{\partial \boldsymbol{\zeta}_i \partial \boldsymbol{\zeta}_{j}} 
\left[ {\cal B}_{ij} (t) \Pi \right]\,,
\label{FPE}
\end{equation} 
where we have introduced the notation 
$\boldsymbol{\zeta}_{i} = (\xi_{i},\eta_{i})$. The function 
${\cal A}_{i}(\boldsymbol{\zeta})$ and the matrix ${\cal B}_{ij}$ are given by
\begin{eqnarray}
{\cal A}^{\rm loc}_{i,1} &=& \frac{1}{\Omega}\left[ 2 p_{1}\psi_{i} 
- d_{1} \right]\xi_{i} + \frac{1}{\Omega}\left[ 2 p_{1} \phi_{i} \right] 
\eta_{i}\,, \nonumber \\
{\cal A}^{\rm loc}_{i,2} &=& \frac{1}{\Omega}\left[ - 2\left( p_{1}+p_{2}+
b \right) \psi_{i} \right] \xi_{i} \nonumber \\
&+& \frac{1}{\Omega}\left[ -2\left( p_{1}+p_{2}+b \right) \phi_{i} + 
\left( 2b-d_{2} \right) - 4b\psi_{i} \right] \eta_{i}\,, \nonumber \\
\label{drift_loc}
\end{eqnarray}
and
\begin{eqnarray}
{\cal B}_{ij,11}^{\rm loc} &=& \frac{1}{\Omega}\left(d_{1}\phi_{i}+2p_{1}
\psi_{i}\phi_{i}\right) \delta_{ij}\,, \nonumber \\
{\cal B}_{ij,22}^{\rm loc} &=& \frac{1}{\Omega}\left(2b\psi_{i}
\left(1-\phi_{i}-\psi_{i}\right)+d_{2}\psi_{i} \right. \nonumber \\
&+& 2\left. \left(p_{1}+p_{2}\right)\psi_{i}\phi_{i}\right) \delta_{ij}\,, 
\nonumber \\
{\cal B}_{ij,12}^{\rm loc} &=& {\cal B}_{ij,21}^{\rm loc}=\frac{1}{\Omega}
\left(-2p_{1}\phi_{i}\psi_{i}\right) \delta_{ij}\,. 
\label{coreac1}
\end{eqnarray}
The superscript loc denotes their origin from the local reaction contribution 
of the master equation, and the subscripts $1$ and $2$ refer to 
$\zeta_{1}=\xi$ and $\zeta_{2}=\eta$, respectively. These results agree with
the non-spatial results found in \cite{AJMaTJN05}, up to a factor of 
$\Omega$, as required. It should also be noted that the function 
${\cal A}_{i}(\boldsymbol{\zeta})$ is linear in $\xi_{i}$ and $\eta_{i}$ with 
coefficients which are exactly those which would be obtained from a linear 
stability analysis of Eqs.~(\ref{local_macro_1}) and 
(\ref{local_macro_2})\,\cite{NGvK92}. This is given in the main text by 
Eq.~(\ref{homo_pert}), which agrees with the results in Eq.~(\ref{drift_loc}).
By contrast the ${\cal B}_{ij}$ cannot be obtained from the macroscopic 
results. 

Next we carry out the same procedures on the contribution due to migration,
${\cal T}^{\rm mig}_{i}$. To do this, the operator expressions listed below 
are required:
\begin{eqnarray}
E_{x_{i}}^{-1}E_{x_{j}}-1&=&N^{-\frac{1}{2}}\left[\frac{\partial}
{\partial\xi_{j}}-\frac{\partial}{\partial\xi_{i}}\right]\nonumber\\
&+&\frac{1}{2}N^{-1}\left[\frac{\partial}{\partial\xi_{i}}
-\frac{\partial}{\partial\xi_{j}}\right]^2\,,\nonumber\\
E_{x_{i}}E_{x_{j}}^{-1}-1&=&N^{-\frac{1}{2}}\left[\frac{\partial}
{\partial\xi_{i}}-\frac{\partial}{\partial\xi_{j}}\right]\nonumber\\
&+&\frac{1}{2}N^{-1}\left[\frac{\partial}{\partial\xi_{i}}
-\frac{\partial}{\partial\xi_{j}}\right]^2,\nonumber\\
E_{y_{i}}^{-1}E_{y_{j}}-1&=&N^{-\frac{1}{2}}\left[\frac{\partial}
{\partial\eta_{j}}-\frac{\partial}{\partial\eta_{i}}\right]\nonumber\\
&+&\frac{1}{2}N^{-1}\left[\frac{\partial}{\partial\eta_{i}}
-\frac{\partial}{\partial\eta_{j}}\right]^2,\nonumber\\
E_{y_{i}}E_{y_{j}}^{-1}-1&=&N^{-\frac{1}{2}}\left[\frac{\partial}
{\partial\eta_{i}}-\frac{\partial}{\partial\eta_{j}}\right]\nonumber\\
&+&\frac{1}{2}N^{-1}\left[\frac{\partial}{\partial\eta_{i}}
-\frac{\partial}{\partial\eta_{j}}\right]^2\,.
\label{E-2}
\end{eqnarray}
These operators possess the general structure 
$N^{-\frac{1}{2}}\hat{L}_{1}+N^{-1}\hat{L}_{2}$, with $\hat{L}_{1}$ equal
to a difference of first derivatives and $\hat{L}_{2}= \hat{L}^{2}_{1}/2$. 
In addition the transition rates (\ref{nn}) have a common structure as 
functions of $N$ which is 
$\rho\left(NF_{1}+N^{\frac{1}{2}}F_{2}+F_{3}+\ldots\right)$, when 
written in terms of the new variables, with $\rho=\mu_{1}/(z\Omega)$ 
or $\mu_{2}/(z\Omega)$, depending on which term one is considering. The
$F_k$ depend on the macroscopic fractions ($\phi_{i}$ and $\psi_{i}$) and on
the stochastic variables ($\boldsymbol{\zeta}_i$), except for $F_1$ which only 
depends on the former. Therefore the form of the part of the master equation 
involving migration terms is
\begin{eqnarray}
& &\left[N^{-\frac{1}{2}}\hat{L}_{1}+N^{-1}\hat{L}_{2}\right]
\rho\left(NF_{1}+N^{\frac{1}{2}}F_{2}+F_{3}\right) \Pi\nonumber\\
&=&\rho\left[N^{\frac{1}{2}}F_{1}\hat{L}_{1}+\hat{L}_{1}F_{2}+
F_{1} \hat{L}_{2}+\dots\right]\Pi\,,
\label{diffgscheme}
\end{eqnarray}
keeping only terms of the order required. This allows us to identify the three
main contributions:
\begin{itemize}
\item[(a)] The order $N^{\frac{1}{2}}$ term is identified with the second term
in the left-hand side of the master equation (Eq.~(\ref{LHS}) with 
$\dot{\xi}_{i}=-(N)^{\frac{1}{2}}\dot{\phi}_{i}$ and 
$\dot{\eta}_{i}=-(N)^{\frac{1}{2}}\dot{\psi}_{i}$) which leads to $2\Omega$ 
independent macroscopic equations.

\item[(b)] The order $N^{0}$ term $\rho\hat{L}_{1}F_{2}$ is of the same 
order as the time-derivative in Eq.~(\ref{LHS}). Since it involves only
first-order derivatives in $\boldsymbol{\zeta}_{i}$ it will give contributions 
which will add to the ${\cal A}_{i}$ in Eq.~(\ref{FPE}) found for the purely 
local terms in the master equation.

\item[(c)] The order $N^{0}$ term $\rho F_{1} \hat{L}_{2}$ is also of the 
same order as the time-derivative in Eq.~(\ref{LHS}). Since it involves only
second-order derivatives in $\boldsymbol{\zeta}_{i}$ it will give contributions
which will add to the ${\cal B}_{ij}$ in Eq.~(\ref{FPE}) found for the purely 
local terms in the master equation.
\end{itemize}

As an example, the term $T_{n_{i}+1,n_{j}-1|n_{i},n_{j}}$ in Eq.~(\ref{nn}) 
when written out in the new variables gives
\begin{eqnarray}
& &\frac{\mu_{1}}{z\Omega}\left[\left\{ \phi_{j}\left(1-\phi_{i}-
\psi_{i}\right)\right\}N+\left\{\left(1-\phi_{i}-\psi_{i}\right)
\xi_{j}\right.\right.\nonumber\\
&-&\left.\left.\phi_{j}\left(\xi_{i}+\eta_{i}\right)\right\}N^{\frac{1}{2}}-
\xi_{j}\left(\xi_{i}+\eta_{i}\right)\right] \Pi\,.
\label{first_T_mig}
\end{eqnarray}
In the notation we have introduced above
\begin{equation}
F_{1} = \phi_{j}\left(1-\phi_{i}-\psi_{i}\right)\,.
\label{F_1_1}
\end{equation}
The second term in Eq.~(\ref{nn}), $T_{n_{i}-1,n_{j}+1|n_{i},n_{j}}$, can be 
obtained from the first term by interchanging $i$ and $j$ (and this is still
true when the operators are included in Eq.~(\ref{mig_contrib})), so adding 
these expression together we find 
\begin{equation}
-\frac{2\mu_{1}}{z\Omega}\left[\sum_{j}\left(\phi_{j}-\phi_{i}\right)
+\sum_{j}\left(\phi_{i}\psi_{j}-\phi_{j}\psi_{i}\right)\right]\,.
\end{equation}
To obtain this we have identified $\partial \Pi/\partial \xi_{i}$, for each 
$i$, with the corresponding term on the left-hand side of the master 
equation (\ref{LHS}). Using the discrete Laplacian operator
\begin{equation}
\Delta f_{i}=\frac{2}{z}\sum_{j \in i}\left(f_{j}-f_{i}\right)\,,
\label{discrete_Laplacian}
\end{equation}
this may be written as
\begin{equation}
-\frac{\mu_{1}}{\Omega}\left[\Delta\phi_{i}+\phi_{i}\Delta\psi_{i}
-\psi_{i}\Delta\phi_{i}\right]\,.
\label{CrossMacro1}
\end{equation}
A similar analysis may be carried out for the terms
\[
\left(E_{y_{i}}^{-1}E_{y_{j}}-1\right)T_{m_{i}+1,m_{j}-1|m_{i},m_{j},}\]
and \[
\left(E_{y_{i}}E_{y_{j}}^{-1}-1\right)T_{m_{i}-1,m_{j}+1|m_{i},m_{j}}.\]
This will give the same form as above, but with the obvious changes 
$\mu_{1} \to \mu_{2}$, $\psi_{i} \leftrightarrow \phi_{i}$, etc.. For the 
macroscopic contribution one thus finds
\begin{equation}
-\frac{\mu_{2}}{\Omega}\left[\Delta\psi_{i}+\psi_{i}\Delta\phi_{i}
-\phi_{i}\Delta\psi_{i}\right]\,.
\label{CrossMacro2}
\end{equation}
Adding Eq.~(\ref{CrossMacro1}) to the right-hand side of 
Eq.~(\ref{local_macro_1}) and Eq.~(\ref{CrossMacro2}) to the right-hand side 
of Eq.~(\ref{local_macro_2}) gives the set of macroscopic laws 
Eqs.~(\ref{dl1})-(\ref{dl2}) for each patch $i$. 

Returning to the stochastic contributions, the one of type (b) coming 
from the term\[
\left(E_{x_{i}}^{-1}E_{x_{j}}-1\right)T_{n_{i}+1,n_{j}-1|n_{i},n_{j}}\,,\]
is the $F_2$-type term in Eq.~(\ref{first_T_mig}). Explicitly this is equal to
\begin{equation}
\frac{\mu_1}{z\Omega} \sum_{i,j} \left[\frac{\partial}
{\partial\xi_{j}}-\frac{\partial}{\partial\xi_{i}}\right]
\left[ \left(1-\phi_{i}-\psi_{i}\right)\xi_{j}-\phi_{j}
\left(\xi_{i}+\eta_{i}\right)\right] \Pi\,.
\end{equation}
The term\[
\left(E_{x_{i}}E_{x_{j}}^{-1}-1\right)T_{n_{i}-1,n_{j}+1|n_{i},n_{j}}\,,\]
gives precisely the same contribution, and adding these together one finds
\begin{eqnarray}
& & - \frac{\mu_{1}}{\Omega}\sum_{i} \frac{\partial}{\partial \xi_i}
\left[ \left\{ \Delta - \psi_{i}\Delta + 
\left(\Delta\psi_{i}\right) \right\}\xi_{i} \right. \nonumber \\
& & + \left. \left\{ \phi_{i} \Delta - 
\left(\Delta \phi_{i}\right) \right\} \eta_{i} \right] \Pi\,.
\label{DiffLang1}
\end{eqnarray}
This may be written as
\begin{equation}
- \frac{\mu_{1}}{\Omega}\sum_{i} \frac{\partial}{\partial \xi_i}
\left[ D_{i, 11}\,\xi_{i} + D_{i, 12}\,\eta_{i} \right] \Pi\,,
\label{mig_type(b)_1}
\end{equation}
where
\begin{equation}
D_{i, 11} = \Delta - \psi_{i}\Delta + \left(\Delta\psi_{i}\right)\,, \ \
D_{i, 12} = \phi_{i} \Delta - \left(\Delta \phi_{i}\right)\,.
\label{D_1}
\end{equation}
In an analogous way, the migrational contributions from the third and fourth 
terms in Eq.~(\ref{nn}) give (letting $\mu_{1} \to \mu_{2}$, 
$\phi_{i} \leftrightarrow \psi_{i}$ and $ \xi_{i} \leftrightarrow \eta_{i}$)
\begin{equation}
- \frac{\mu_{2}}{\Omega}\sum_{i} \frac{\partial}{\partial \eta_i}
\left[ D_{i, 21}\,\xi_{i} + D_{i, 22}\,\eta_{i} \right] \Pi\,,
\label{mig_type(b)_2}
\end{equation}
where
\begin{equation}
D_{i, 22} = \Delta - \phi_{i}\Delta + \left(\Delta\phi_{i}\right)\,, \ \
D_{i, 21} = \psi_{i} \Delta - \left(\Delta \psi_{i}\right)\,. 
\label{D_2}
\end{equation}
The results (\ref{mig_type(b)_1})-(\ref{D_2}) can also be obtained through a
linear-stability analysis of the non-local terms in 
Eqs.~(\ref{dl1})-(\ref{dl2}). They represent diffusion and should be added  
to the terms in Eq.~(\ref{drift_loc}) which represent reactions, to give the 
complete contribution in the first term on the right-hand side of the 
Fokker-Planck equation (\ref{FPE}). 

Finally, there are the terms of type (c), which have the form 
$\rho F_{1} \hat{L}_{2}$. We have already discussed the $F_{1}$ terms, and the
operators $\hat{L}_{2}$ may be read off from Eq.~(\ref{E-2}). The four terms
corresponding to those in Eq.~(\ref{nn}) are:
\begin{equation}
\begin{array}{c}
\frac{\mu_{1}}{z\Omega}\sum_{i,j}\frac{1}{2}\left[\phi_{i}\left(1-\phi_{j}
-\psi_{j}\right)\right]\left[\frac{\partial}{\partial\xi_{i}}
-\frac{\partial}{\partial\xi_{j}}\right]^{2} \Pi\,,\\
\frac{\mu_{1}}{z\Omega}\sum_{i,j}\frac{1}{2}\left[\phi_{j}\left(1-\phi_{i}
-\psi_{i}\right)\right]\left[\frac{\partial}{\partial\xi_{i}}
-\frac{\partial}{\partial\xi_{j}}\right]^{2} \Pi\,,\\
\frac{\mu_{2}}{z\Omega}\sum_{i,j}\frac{1}{2}\left[\psi_{i}\left(1-\phi_{j}
-\psi_{j}\right)\right]\left[\frac{\partial}{\partial\eta_{i}}-
\frac{\partial}{\partial\eta_{j}}\right]^{2} \Pi\,,\\
\frac{\mu_{2}}{z\Omega}\sum_{i,j}\frac{1}{2}\left[\psi_{j}\left(1-\phi_{i}
-\psi_{i}\right)\right]\left[\frac{\partial}{\partial\eta_{i}}
-\frac{\partial}{\partial\eta_{j}}\right]^{2} \Pi\,.
\end{array}
\label{Spacorrelators}
\end{equation}
In this paper we will only be interested in studying the equations satisfied 
by the stochastic variables $\boldsymbol{\zeta}_{i}=(\xi_{i}, \eta_{i})$,\
$i=1,\ldots,\Omega$, when the transients in the macroscopic equations
(\ref{dl1})-(\ref{dl2}) have died away. Then $\phi_{i}$ and $\psi_{i}$ are
equal to their fixed point values $\phi^*$ and $\psi^*$ respectively, which
are independent of the site label $i$. Adding the four 
contributions (\ref{Spacorrelators}) in this case gives
\begin{eqnarray}
& & \frac{2\mu_{1}}{z\Omega} \phi^{*} \left( 1 - \phi^{*} - \psi^{*} \right)
\sum_{i,j} \left[ z \delta_{ij} \frac{\partial^2}{\partial\xi^2_{i}}
-\frac{\partial^2}{\partial\xi_{j}\partial\xi_{j}}\right] \Pi \nonumber \\
&+& \frac{2\mu_{2}}{z\Omega} \psi^{*} \left( 1 - \phi^{*} - \psi^{*} \right)
\sum_{i,j} \left[ z \delta_{ij} \frac{\partial^2}{\partial\eta^2_{i}}
-\frac{\partial^2}{\partial\eta_{j}\partial\eta_{j}}\right] \Pi\,. 
\nonumber \\ 
\label{B-mig}
\end{eqnarray}
These contributions are diagonal in the predator-prey variables (there are no
mixed derivatives involving $\xi$ and $\eta$), but is not diagonal in the
site variables (there are mixed derivatives involving $i$ and $j$). Comparing
Eq.~(\ref{B-mig}) with the Fokker-Planck equation (\ref{FPE}), we see that 
the contributions to the matrix ${\cal B}$, which add to those in 
Eq.~(\ref{coreac1}) are
\begin{eqnarray}
{\cal B}^{\rm mig}_{ij, 11} &=& \frac{4\mu_{1}}{\Omega} \phi^{*} 
\left( 1 - \phi^{*} - \psi^{*} \right) \delta_{ij} \nonumber \\
&-& \frac{4\mu_{1}}{z\Omega} \phi^{*} 
\left( 1 - \phi^{*} - \psi^{*} \right) J_{\langle ij \rangle}\,, \nonumber \\
{\cal B}^{\rm mig}_{ij, 22} &=& \frac{4\mu_{2}}{\Omega} \psi^{*} 
\left( 1 - \phi^{*} - \psi^{*} \right) \delta_{ij} \nonumber \\
&-& \frac{4\mu_{2}}{z\Omega} \psi^{*} 
\left( 1 - \phi^{*} - \psi^{*} \right) J_{ \langle ij \rangle}\,, 
\label{coreac2}
\end{eqnarray}
where $J_{\langle ij \rangle}$ is zero unless $i$ and $j$ are nearest 
neighbors.

In summary, the order $N^{0}$ terms give the Fokker-Planck equation 
(\ref{FPE}), with the function ${\cal A}_{i}(\boldsymbol{\zeta})$ and the 
matrix ${\cal B}_{ij}$ being given by:
\begin{eqnarray}
{\cal A}_{i,1} &=& \alpha_{i,11}\xi_{i} + \alpha_{i,12}\eta_{i} \nonumber \\
{\cal A}_{i,2} &=& \alpha_{i,21}\xi_{i} + \alpha_{i,22}\eta_{i}\,,
\label{A_def}
\end{eqnarray}
where the $\alpha$ are exactly the coefficients found in Section \ref{Turing}
by linear stability analysis, and 
\begin{eqnarray}
{\cal B}_{ij,11} &=& \left[ \left(d_{1}\phi^{*}+2p_{1} \psi^{*}\phi^{*}\right) 
+ 4\mu_{1} \phi^{*} \left( 1 - \phi^{*} - \psi^{*} \right) \right] 
\delta_{ij} \nonumber \\
&-& \frac{4\mu_{1}}{z} \phi^{*} 
\left( 1 - \phi^{*} - \psi^{*} \right) J_{\langle ij \rangle}\,, \nonumber \\
{\cal B}_{ij,22} &=& \left[ \left(2b\psi^{*} \left(1-\phi^{*}-\psi^{*}\right)+
d_{2}\psi^{*} \right. \right. \nonumber \\
&+& 2\left. \left. \left(p_{1}+p_{2}\right)\psi^{*}\phi^{*}\right) 
+ 4\mu_{2} \psi^{*} \left( 1 - \phi^{*} 
- \psi^{*} \right) \right] \delta_{ij} \nonumber \\
&-& \frac{4\mu_{2}}{z} \psi^{*} \left( 1 - \phi^{*} - \psi^{*} \right) 
J_{ \langle ij \rangle}\,, \nonumber \\
{\cal B}_{ij,12} &=& {\cal B}_{ij,21} = \left[ -2p_{1} \phi^{*}\psi^{*}\right] 
\delta_{ij}\,. 
\label{B_def}
\end{eqnarray}
In the above we have assumed that the Fokker-Planck equation (\ref{FPE}) has
been re-expressed in terms of the rescaled time $\tau = t/\Omega$, in order
to eliminate factors of $\Omega^{-1}$ from ${\cal A}$ and ${\cal B}$. 

\section{Fourier analysis}
\label{FT}

As discussed in the main text we carry out a temporal Fourier transform in
order to calculate the power spectra associated with the fluctuations about 
the stationary state in order to identify temporal cycles, but we also wish 
to carry out spatial Fourier transforms. There are a number of reasons for
doing this: (a) the translational invariance of the stationary state means
that quantities of interest become diagonal in Fourier space, (b) because of
this the continuum limit is easily taken, and (c) the power spectra are 
naturally generalized from the non-spatial case to depend on the wave-vector 
as well as on the frequency.

We largely follow the conventions of Chaitin and Lubensky\,\cite{cha95} in 
introducing the spatial Fourier transforms. That is, we define the Fourier 
transform, $f_{\bf k}$, of a function $f_{\bf j}$ defined on a $d-$dimensional 
hypercubic lattice, with lattice spacing $a$, by
\begin{eqnarray}
f_{\bf k} &=& a^{d} \sum_{\bf j} e^{ - i {\bf k}.a{\bf j}}\,f_{\bf j}\,,
\nonumber \\
f_{\bf j} &=& a^{-d}\,\Omega^{-1}\,\sum_{\bf k} e^{ i {\bf k}.a{\bf j}}
\,f_{\bf k}\,,
\label{FTdef}
\end{eqnarray}
where, for clarity, we have deviated from the usual notation of the main 
text and written the lattice site label ${\bf j}$ as a vector. Here ${\bf k}$
is restricted to the first Brillouin zone: 
$-(\pi/a) \leq k_{\gamma} \leq (\pi/a)$, $\gamma=1,\ldots,d$. We will also
require the result\,\cite{cha95}
\begin{equation}
\sum_{\bf j} e^{ - i {\bf k}.a{\bf j}} = \Omega\,\delta_{{\bf j},0}\,.
\label{FTres}
\end{equation}

Using the definition (\ref{FTdef}) we may take the Fourier transform of the
Langevin equation (\ref{Langevin}). This is straightforward for the time
derivative on the left-hand side and for the noise term 
$\boldsymbol{\lambda}_i$. For the ${\cal A}_i$ term we use Eq.~(\ref{A_def})
where the $\alpha$ are made up of the local constant terms (\ref{homo_pert})
and those coming from diffusion (\ref{D_1}) and (\ref{D_2}). At the fixed 
point where $\phi$ and $\psi$ are homogeneous these diffusion operators 
are site-independent and given by $D_{11}=(1-\psi^{*})\Delta$, 
$D_{12}=\phi^{*}\Delta$, $D_{21}=\psi^{*}\Delta$ and 
$D_{22}=(1-\phi^{*})\Delta$. The Fourier transform of the Langevin equation 
thus takes the form (\ref{Lang}), with the $\alpha$ given by 
Eq.~(\ref{alphas}), where $\Delta_{\bf k}$ is the Fourier transform of the 
discrete Laplacian operator $\Delta$. From the definitions 
(\ref{discrete_Laplacian}) and (\ref{FTdef}) this is easily shown to be given 
by Eq.~(\ref{dis_Lap_k}).

To complete the description of the Langevin equation in ${\bf k}-$space, we
need rewrite the correlation function (\ref{correlators}). Taking the 
Fourier transform of both $\boldsymbol{\lambda}_i(\tau)$ and 
$\boldsymbol{\lambda}_j(\tau')$ yields
\begin{equation}
\langle \boldsymbol{\lambda}_{\bf k}(\tau) 
\boldsymbol{\lambda}_{\bf k'}(\tau') \rangle = a^{2d}\sum_{\bf i,j} 
e^{ - i {\bf k}.a{\bf i}}\,e^{ - i {\bf k'}.a{\bf j}}\,{\cal B}_{\bf ij}\,
\delta(\tau-\tau')\,.
\label{FT_corr}
\end{equation}
However, ${\cal B}_{\bf ij}$ is given by Eq.~(\ref{B_def}) and is only 
non-zero if ${\bf i}={\bf j}$ or if ${\bf i}$ and ${\bf j}$ are 
nearest-neighbors. That is, it has the form
\begin{equation}
{\cal B}_{\bf ij} = b^{(0)}\,\delta_{\bf ij} + b^{(1)}\,
J_{\langle {\bf ij} \rangle}\,.
\label{form_of_B}
\end{equation}
The translational invariance of ${\cal B}_{\bf ij}$ is quite clear: it can be
completely specified by the difference ${\bf d} = {\bf j} - {\bf i}$:
\begin{equation}
{\cal B}_{\bf d} = b^{(0)}\,\delta_{{\bf d},0} + b^{(1)}\,
\delta_{|{\bf d}|,1}\,.
\label{second_form_of_B}
\end{equation}
Inserting the expression for ${\cal B}_{\bf d}$ in terms of its Fourier 
transform, ${\cal B}_{\bf q}$, in Eq.~(\ref{FT_corr}), we have from 
Eqs.~(\ref{FTdef}) and (\ref{FTres}) that
\begin{eqnarray}
\langle \boldsymbol{\lambda}_{\bf k}(\tau) 
\boldsymbol{\lambda}_{\bf k'}(\tau') \rangle &=& a^{d}\,\Omega\,
\sum_{\bf q} {\cal B}_{\bf q}\,\delta_{{\bf k},{\bf q}}\,
\delta_{{\bf k}',{\bf -q}}\,\delta(\tau-\tau')\nonumber \\
&=& {\cal B}_{\bf k}\,a^{d}\Omega\,\delta_{{\bf k}+{\bf k}',0}\,
\delta(\tau-\tau')\,.
\label{FT_corr1}
\end{eqnarray}
Now
\begin{eqnarray}
{\cal B}_{\bf k} &=& a^{d}\,\sum_{\bf d} e^{ - i {\bf k}.a{\bf d}}\,
{\cal B}_{\bf d} \nonumber \\
&=& a^{d}\,\left\{ b^{(0)} + 2 b^{(1)} \left[ \sum^{d}_{\gamma = 1} 
\cos \left( k_{\gamma} a \right) \right] \right\}
\label{FT_corr_2}
\end{eqnarray}
using Eq.~(\ref{second_form_of_B}). In terms of $\Delta_{\bf k}$ defined by 
Eq.~(\ref{dis_Lap_k}), this may be written as
\begin{equation}
{\cal B}_{\bf k} = a^{d}\,\left\{ \left[ b^{(0)} + z b^{(1)} \right]
+ \frac{z b^{(1)}}{2}\,\Delta_{\bf k} \right\}\,,
\label{B_k_I}
\end{equation}
since for a hypercubic lattice the coordination number is $z=2d$. Writing these
out explicitly using Eqs.~(\ref{B_def}) and (\ref{form_of_B}) gives 
Eq.~(\ref{B_k}) in the main text.

Finally, we can ask what happens as we take the lattice spacing, $a$,
to zero, but keeping $\Omega a^{d}$ (the area, if $d=2$) fixed. Using 
Eq.~(\ref{dis_Lap_k}) and
\begin{align}
\cos\left(k_{\gamma}a\right) & \simeq1-\frac{\left(k_{\gamma}a\right)^{2}}{2}
+O(\left(ka\right)^{4})
\label{kdef}
\end{align}
we see that $\Delta_{\bf k} = - a^{2} k^{2}/d + O(k^{4})$. Since 
$\Delta_{\bf k}$ always appears along with the migration rates, the factor of
$a^{2}/d$ can always be absorbed into these rates by defining new quantities
\begin{equation}
\tilde{\mu}_{1} = \frac{1}{d} a^{2} \mu_{1}\,, \ \ 
\tilde{\mu}_{2} = \frac{1}{d} a^{2} \mu_{2}\,.
\label{scaled_mig_rates}
\end{equation}
So for instance, in Eqs.~(\ref{alphas}) and (\ref{B_k}) the $\Delta_{\bf k}$ 
can be replaced by $-k^{2}$ and $\mu_1$ and $\mu_2$ by $\tilde{\mu}_1$ and
$\tilde{\mu}_2$ respectively, as $a$ becomes small (or equivalently $\Omega$
becomes large). In this limit $\Omega\,a^{d}\,\delta_{{\bf k}+{\bf k}', 0}$ 
becomes $(2\pi)^{d}\,\delta({\bf k}+{\bf k}')$\,\cite{cha95}, and therefore
Eq.~(\ref{corr}) becomes
\begin{equation}
\langle {\bf{\lambda}}_{\bf k} (\tau) 
{\bf{\lambda}}_{{\bf k}'} (\tau') \rangle = {\cal B}_{\bf k}\,(2\pi)^{d}
\delta({\bf k}+{\bf k}')\,\delta(\tau-\tau')\,,
\label{corr_cont}
\end{equation}
where $B_{\bf k}$ is given by Eq.~(\ref{B_k}), but with the small $a$ 
approximation described above.

To obtain the power spectrum we need to take the temporal Fourier transform
of Eq.~(\ref{corr_cont}). This yields
\begin{equation}
\left\langle \boldsymbol{\lambda}_{\bf k} (\omega)
\boldsymbol{\lambda}_{{\bf k}'} (\omega')\right\rangle =
{\cal B}_{\bf k} \left(2\pi\right)^{d}
\delta\left({\bf k}+{\bf k}'\right)\,\left(2\pi\right)
\delta\left(\omega+\omega'\right)\,.
\label{full_FT}
\end{equation}
Since there are only contributions in the above formula when 
${\bf k}'=-{\bf k}$ and $\omega'=-\omega$ this is frequently written as 
\begin{equation}
\left\langle \boldsymbol{\lambda}_{\bf k} (\omega)
\boldsymbol{\lambda}_{-{\bf k}} (-\omega) \right\rangle =
{\cal B}_{\bf k}\,,
\label{abbrev_FT}
\end{equation}
or equivalently, since 
$\boldsymbol{\lambda}^{*}_{\bf k} (\omega)= 
\boldsymbol{\lambda}_{-{\bf k}} (-\omega)$, as in Eq.~(\ref{double_corr}).

\end{document}